\documentclass[12pt]{article}
\usepackage{amssymb,amsmath,epsfig}
\allowdisplaybreaks

\begin{document}

\title{\bf Study of Static Wormhole Solutions in $F(T,T_\mathcal{G})$ Gravity}

\author{M. Sharif \thanks{msharif.math@pu.edu.pk}~ and
Kanwal Nazir \thanks{awankanwal@yahoo.com}~~\thanks{On leave from
Department of Mathematics, Lahore College
for Women University, Lahore-54000, Pakistan.}\\
Department of Mathematics, University of the Punjab,\\
Quaid-e-Azam Campus, Lahore-54590, Pakistan.}

\date{}
\maketitle
\begin{abstract}
In this paper, we investigate static spherically symmetric wormhole
solutions in the background of $F(T,T_\mathcal{G})$ gravity ($T$ is
the torsion scalar and $T_{\mathcal{G}}$ represents teleparallel
equivalent of the Gauss-Bonnet term). We study the wormhole
solutions by assuming four different matter contents, a specific
redshift function and a particular $F(T,T_\mathcal{G})$ model. The
behavior of null/weak energy conditions for these fluids is analyzed
graphically. It turns out that wormhole solutions can be obtained in
the absence of exotic matter for some particular regions of
spacetime. We also explore stability of wormhole solutions through
equilibrium condition. It is concluded that there exist physically
acceptable wormhole solutions for anisotropic, isotropic and
traceless fluids.
\end{abstract}

\section{Introduction}

The current accelerated expanding behavior of the universe is
confirmed through several cosmological observations. The rapid rate
of expansion indicates the presence of an anonymous force other than
dark matter and baryonic matter in the universe. This mysterious
force is labeled as dark energy (DE) which is equally scattered in
the universe with negative pressure. Its mysterious nature can be
explained through two renowned proposals. The first modifies the
matter part while the second establishes the gravitational
modification of the Einstein-Hilbert action leading to modified
theories such as Gauss-Bonnet (GB) theory \cite{1}, $f(R)$ theory
($R$ represents the Ricci scalar) \cite{2}, $F(T)$ theory \cite{3},
$f(R,\mathcal{T})$ theory ($\mathcal{T}$ defines trace of the
energy-momentum tensor) \cite{4} etc. Recently, another modification
is introduced by incorporating $T$ and $T_\mathcal{G}$ known as
$F(T,T_\mathcal{G})$ gravity.

The $F(T,T_\mathcal{G})$ theory is an extension of $F(T)$ theory
which is obtained by inserting the higher-order torsion invariants
in the action. This is a torsion based modification with no
curvature formulation. The motivation behind this extension is that
in curvature based theory such as $f(R)$ theory, higher-order
curvature corrections like GB term $\mathcal{G}$ and function
$f(\mathcal{G})$ are introduced in the action. On the same pattern,
one can construct torsion based theory by involving higher-order
torsion corrections terms in the action. A lot of work has been done
to study different cosmological features using this theory
\cite{6}-\cite{7}. The study of different matter contents is of
great interest in modified theories. These matter distributions are
helpful in explaining the matter contents of the astronomical
objects like wormholes.

A wormhole is known as a hypothetical path like a tunnel or bridge
that provides a connection between two different regions of the
universe apart from one another. The existence of a realistic
wormhole which satisfies the energy conditions has always been a
challenging issue. Dynamical wormhole solutions \cite{13},
traversable wormholes \cite{13a}, brane wormholes \cite{12},
generalized Chaplygin gas \cite{14} etc are used to minimize the
violation of energy conditions especially null energy condition
(NEC). Rahaman et al. \cite{19} studied wormhole solutions by
considering noncommutative geometry and noticed the presence of
asymptotically flat solutions for four dimensions. Abreu and Sasaki
\cite{20} explored the effects of energy conditions through
noncommutative wormhole in the absence of exotic matter.

A comprehensive study of wormhole solutions has been done in
modified theories. Lobo and Oliveira \cite{17} studied wormhole
solutions with different fluids in $f(R)$ theory and explored energy
conditions. B$\ddot{o}$hmer et al. \cite{18} considered a particular
model to derive traversable wormhole solutions in $F(T)$ gravity and
found that there exists a physically acceptable wormhole. Sharif and
Rani \cite{21} assumed a particular shape function as well as $F(T)$
model with noncommutative geometry and found that energy conditions
violate due to the presence of effective energy-momentum tensor but
noncommutative geometry does not play any role in this violation. We
discussed wormhole solutions with noncommutative geometry in
$F(T,T_\mathcal{G})$ gravity and concluded that effective
energy-momentum tensor is responsible for the violation of energy
conditions \cite{21a}.

Sharif and Rani \cite{23} explored dynamical wormhole solutions in
the same gravity using anisotropic matter distribution. Sharif and
Zahra \cite{24b} discussed some specific solutions in $f(R)$ gravity
with isotropic, anisotropic and barotropic fluids and found that
physically acceptable wormholes exist only for barotropic fluid in
some particular regions. Sharif and Ikram \cite{24a} explored energy
conditions for static wormhole solutions with same fluids in
$F(\mathcal{G})$ gravity. Zubair et al. \cite{24c} assumed the same
fluids in $F(R,\mathcal{T})$ gravity and found that realistic and
stable wormhole solutions exist only for anisotropic case.

In this paper, we study static wormhole solutions in
$F(T,T_\mathcal{G})$ gravity with four matter contents. The paper is
arranged as follows. In section \textbf{2}, we provide necessary
formalism of wormhole geometry as well as energy conditions in this
theory. Section \textbf{3} explores the structure and existence of
the wormhole through shape function, NEC and weak energy condition
(WEC) for four types of fluids and a specific $F(T,T_\mathcal{G})$
model. In section \textbf{4}, we analyze stability of the wormhole
solutions through equilibrium condition. Finally, we conclude our
results.

\section{Formalism of $F(T,T_\mathcal{G})$ Theory}

In this section, we formulate the field equations in the framework
of $F(T,T_\mathcal{G})$ gravity and provide an overview of the
energy conditions as well as wormhole geometry.

\subsection{$F(T,T_\mathcal{G})$ Gravity}

The tetrad field $e_{a}(x^{\mu})$ has a fundamental role in $F(T)$
as well as $F(T,T_\mathcal{G})$ gravity. Trivial tetrad is the
simplest one expressed as $e_{a}=\partial_{\mu}{\delta^{\mu}}_{a}$
and $e^{b}=\partial^{\mu}{\delta_{\mu}}^{b}$, where
$\delta^{\mu}_{a}$ is the Kronecker delta. These are not commonly
used because they provide zero torsion. The non-trivial tetrad have
different behavior, so they are more supportive in describing
teleparallel theory. These tetrad can be represented as
\begin{equation}\nonumber
h_{a}=\partial_{\mu}{h_{a}^{~\mu}},\quad
h^{b}=dx^{\mu}{h^{b}_{~\mu}},
\end{equation}
satisfying
\begin{equation}\nonumber
h_{~\mu}^{a}h^{~\mu}_{b}=\delta^{a}_{b},\quad
h_{~\mu}^{a}h^{~\nu}_{a}=\delta^{\nu}_{\mu }.
\end{equation}
The metric tensor can also be expressed in the product of tetrad
fields as
\begin{equation}\nonumber
g_{\mu\nu}=\eta_{ab}h_{\mu}^{a}h^{b}_{\nu},
\end{equation}
where $\eta_{ab}$~=~diag$(1,-1,-1,-1)$ is the Minkowski metric. The
coordinates on manifold are represented by Greek indices $(\mu
,\nu,...)$ while coordinates on tangent space are characterized by
Latin indices $(a, b,...)$.

The Weitzenb$\ddot{o}$ck connection ${\omega^{a}}_{b}(x^{\mu})$ that
describes parallel transportation, has the following form
\begin{equation}\nonumber
\omega^{c}_{ab}={h^{c}}_{\mu}{h^{\mu}}_{a, b}.
\end{equation}
The structure coefficients $\mathcal{C}^{c}_{ab}$ are defined as
\begin{equation}\nonumber
[h_{a},h_{b}]=h_{c}\mathcal{C}^{c}_{ab},
\end{equation}
where
\begin{equation}\nonumber
\mathcal{C}^{c}_{ab}={h^{\nu}}_{b}{h^{\mu}}_{a}
({h^{c}}_{\mu,\nu}-{h^{c}}_{\nu,\mu}).
\end{equation}
Similarly, we can express the torsion as well as curvature tensors
as
\begin{eqnarray}\nonumber
T^{a}_{bc}&=&-\omega^{a}_{bc}+\omega^{a}_{cb}-\mathcal{C}^{a}_{bc},\\\nonumber
R^{a}_{bcd}&=&-\omega^{e}_{bc}
\omega^{a}_{ed}+\omega^{a}_{bd,c}+\omega^{e}_{bd}\omega^{a}
_{ec}-\mathcal{C}^{e}_{cd}\omega^{a}_{be}-\omega^{a}_{bc,d}.
\end{eqnarray}
The contorsion tensor is defined by
\begin{equation}\nonumber
\mathcal{K}_{abc}=\frac{1}{2}(-T_{bca}-T_{abc}+T_{cab})=-\mathcal{K}_{bac}.
\end{equation}
Finally, the torsion scalars $T$ and $T_\mathcal{G}$ take the form
\begin{eqnarray}\nonumber
T&=&\frac{1}{4}T^{abc}T_{abc}-T_{ab}^{~~a}T^{cb}_{~~c}
+\frac{1}{2}T^{abc}T_{cba},\\\nonumber
T_\mathcal{G}&=&(2{{\mathcal{K}^{a_{3}}}_{eb}\mathcal{K}^{a_{1}a_{2}}}_{a}
{\mathcal{K}^{ea_{4}}}_{f}{\mathcal{K}^{f}}_{cd}
+{\mathcal{K}^{a_{2}}}_{b}{\mathcal{K}^{a_{1}}}_{ea}
{\mathcal{K}^{a_{3}}}_{fc}{\mathcal{K}^{fa_{4}}}_{d}
+2{\mathcal{K}^{a_{3}}}_{eb}
\\\nonumber&\times&{\mathcal{K}^{a_{1}a_{2}}}_{a}
{\mathcal{K}^{ea_{4}}}_{c,d}-2{{\mathcal{K}^{a_{3}}}_{eb}
\mathcal{K}^{a_{1}a_{2}}}_{a}
{\mathcal{K}^{e}}_{fc}{\mathcal{K}^{fa_{4}}}_{d})
\delta^{abcd}_{a_{1}a_{2}a_{3}a_{4}},
\end{eqnarray}
where
$\delta^{abcd}_{a_{1}a_{2}a_{3}a_{4}}=\epsilon^{abcd}\epsilon_{a_{1}a_{2}a_{3}a_{4}}$
while the antisymmetric symbol $\epsilon_{a_{1}a_{2}a_{3}a_{4}}$ has
$\epsilon_{1234}=1$ and $\epsilon^{1234}=-1$.

Kofinas and Saridakis \cite{6} introduced a different torsion
invariant $T_\mathcal{G}$ to formulate a teleparallel equivalent GB
term in $F(T)$ theory. The Gauss-Bonnet term
$\mathcal{G}=R^2-4R^{ab}R_{ab}+R^{abcd}R_{abcd}$ in terms of
Levi-Civita connection is expressed as
\begin{equation}\nonumber
h\mathcal{\tilde{G}}=\textrm{total~diverg}+hT_{\mathcal{G}},
\end{equation}
where $h=\det(h^a_\nu)$. This equation shows that $T_{\mathcal{G}}$
differs from the GB term only by a total derivative. Also, GB term
is not the total derivative. So, the above equation does not imply
that $T_{\mathcal{G}}=0$. The action in the context of
$F(T,T_\mathcal{G})$ gravity can be defined as
\begin{equation}\nonumber
S=\int
\sqrt{-g}\left[\frac{F(T,T_{\mathcal{G}})}{2\kappa^{2}}+\mathcal{L}_{m}\right]d^4x.
\end{equation}
In the above action, $\kappa^{2}=1$, $\sqrt{-g}=\det(h^a_\nu)$ with
$h^a_\nu$ represents the tetrad, $g$ describes the determinant of
metric coefficients and $\mathcal{L}_{m}$ determines the matter
Lagrangian. The basic entity of this theory is tetrad field. The
metric tensor and all the terms involve in the action are also
expressed in terms of tetrad. So, the field equations can be
obtained by varying the action in terms of tetrad field $h^a_\nu$ as
follows
\begin{eqnarray}\nonumber
&&\mathcal{C}^{b}_{~cd}(H^{dca}+2H^{[ac]d})+(-T_G
F_{T_G}(T,T_G)+F(T,T_G)-T F_{T}(T,T_G))\eta^{ab}\\\nonumber
&+&2(H^{[ba]c}-H^{[kcb]a}+H^{[ac]b})\mathcal{C}_{~dc}^{d}+2(-H^{[cb]a}
+H^{[ac]b}+H^{[ba]c})_{,c}+4H^{[db]c}\\\label{1a}&\times&
\mathcal{C}_{(dc)}^{~~~a}+T^{a}_{~cd}H^{cdb}-\mathcal{H}^{ab}=\kappa^2\mathcal{T}^{ab},
\end{eqnarray}
where
\begin{eqnarray}\nonumber
H^{abc}&=&(\eta^{ac}K^{bd}_{~~d}-K^{bca})F_{T}(T,T_{G})+F_{T_{G}}(T,T_{G})[
(\epsilon^{ab}_{~~lf}K^{d}_{~qr}K^{l}_{~dp}\\\nonumber
&+&2K^{bc}_{~~p}\epsilon^{a}_{~dlf}K^{d}_{~qr}
+K^{il}_{~~p}\epsilon_{qdlf}K^{jd}_{~~r})K^{qf}_{~~t}\epsilon^{kprt}
+\epsilon^{ab}_{~~ld}K^{fd}_{~~p}\epsilon^{cprt}(K^{l}_{~fr,t}
\\\nonumber&-&\frac{1}{2}\mathcal{C}^{q}_{~tr}K^{l}_{~fq})
+\epsilon^{cprt}K^{df}_{~p}\epsilon^{al}_{~~df}(K^{b}_{~kr,t}
-\frac{1}{2}\mathcal{C}^{q}_{~tr}K^{b}_{~lq})]
+\epsilon^{cprt}\epsilon^{a}_{~ldf}\\\nonumber&\times&[F_{T_{G}}(T,T_{G})
K^{bl}_{~~[q}K^{df}_{~~r]}\mathcal{C}^{q}_{~pt}+(K^{bl}_{~p}F_{T_{G}}(T,T_{G})
K^{df}_{~r})_{,t}],\\\nonumber
\mathcal{H}^{ab}&=&F_{T}(T,T_G)\epsilon^{a}_{~lce}K^{l}
_{~fr}\epsilon^{brte}K^{fc}_{~~t},~~
F_{T}(T,T_{\mathcal{G}})=\frac{d}{d T}
F(T,T_{\mathcal{G}}),\\\nonumber
F_{T_{\mathcal{G}}}(T,T_{\mathcal{G}})&=&\frac{d}{d T_{\mathcal{G}}}
F(T,T_{\mathcal{G}}).
\end{eqnarray}
Here, $\mathcal{T}^{ab}$ describes the energy-momentum tensor for
matter field. Notice that we can obtain teleparallel equivalent to
general relativity (GR) for $F(T,T_{G})=-T$ whereas $F(T)$ theory is
achieved for $T_G=0$.

\subsection{Wormhole Geometry}

A wormhole is known as a hypothetical path like a tunnel or bridge
that provides a connection between two different regions of the
universe apart from one another. The existence of a realistic
wormhole which satisfies the energy conditions has always been a
challenging issue. A wormhole through which one can traverse freely
is termed as traversable wormhole. The traversability of the
wormholes is based on the existence of exotic matter in the wormhole
tunnel which violates the NEC. One can easily pass through the
tunnel as exotic matter keeps it open but its enough amount give
rise to non-realistic wormhole. Hence, the physically viable
wormhole solutions exist only if we minimize the amount of this
problematic matter in the wormhole throat.

Morris and Thorne \cite{11} were the first who established the
notion of traversable wormholes by avoiding event horizon. They
proposed a static spherically symmetric metric describing the
wormhole geometry \cite{11} as
\begin{equation}\label{2}
ds^2=e^{2\lambda(r)}dt^2-e^{\chi(r)}dr^2-r^2d\Omega^2,
\end{equation}
where $e^{\chi(r)}=\left(1-\frac{\psi(r)}{r}\right)^{-1}$ and
$d\Omega^2=d\theta^2+\sin^2\theta d\phi^2$. The above metric
entirely depends on two functions, namely redshift function
$\lambda(r)$ and shape function $\psi(r)$. The redshift function
gives gravitational redshift and shape function determines shape of
the wormhole. In order to satisfy the demand of traversable
wormhole, we must have small tidal gravitational forces at the
throat of the wormhole. For the Schwarzschild wormhole, these forces
are so strong that any traveler, who wants to pass through the
throat, would be finished. Consequently, the tidal gravitational
forces, that affect the traveler, must be sufficiently small.

To avoid event horizon as well as strong tidal forces for a
traversable wormhole, we usually assume a non-zero redshift function
which is finite everywhere. Hence we assume the function
$\lambda(r)$ as
\begin{eqnarray}\label{6}
\lambda(r)=-\frac{\zeta}{r},\quad\zeta>0,
\end{eqnarray}
which is non-zero and finite. This satisfies the condition of no
horizon as well as asymptotic flatness. To avoid strong tidal forces
at throat, we have chosen $\zeta$ to be small throughout the paper.

For traversable wormhole, the following properties must be satisfied
by $\psi(r)$ and $\lambda(r)$.
\begin{itemize}
\item The redshift function must fulfill the no horizon
property, i.e., it must be finite throughout.
\item The asymptotic flatness condition ($\frac{\psi(r)}{r}\rightarrow0$ as $r\rightarrow\infty$)
should be an essential constituent of the spacetime at large
distances.
\item The flaring out property $(\frac{\psi(r)-r\psi'(r)}{\psi^2(r)}>0)$ must be satisfied on the
wormhole throat radius $r_{th}$ to obtain an ordinary wormhole
solution. Moreover, $\psi(r)$ satisfies $\psi'(r_{th})<1$ and
$\psi(r)=r_{th}$ at $r=r_{th}$.
\item The condition $\left(1-\frac{\psi(r)}{r}\right)>0$ must be satisfied at the throat.
\end{itemize}

To examine the wormhole solutions, we take a diagonal tetrad
\cite{11} as
\begin{eqnarray}\label{3}
h^{a}_{\nu}=diag\left(e^{-\lambda(r)},\left(1-\frac{\psi(r)}{r}\right)
^{-{\frac{1}{2}}},~r,~r\sin\theta\right).
\end{eqnarray}
This diagonal tetrad is the simplest and frequently used tetrad for
the Morris and Thorne static spherically symmetric metric. The
expressions for the torsion scalars take the following form
\begin{eqnarray}\label{4}
T&=&\frac{2}{r^2}\left(1-\frac{\psi(r)}{r}\right)+\frac{4\lambda'}{r}\left(1-\frac{\psi(r)}{r}\right),\\\nonumber
T_\mathcal{G}&=&\frac{8\psi(r)\lambda'(r)}{r^4}-\frac{8\psi(r)\lambda'^2(r)}{r^3}
\left(1-\frac{\psi(r)}{r}\right)+\frac{12\psi(r)\lambda'(r)\psi'(r)}{r^4}\\\label{5}
&-&\frac{8\psi'(r)\lambda'(r)}{r^3}
-\frac{12\psi^2(r)\lambda'(r)}{r^5}-\frac{8\psi(r)\lambda''(r)}{r^3}
\left(1-\frac{\psi(r)}{r}\right).
\end{eqnarray}
For anisotropic distribution, we assume the following
energy-momentum tensor
\begin{eqnarray}\nonumber
\mathcal{T}^{(m)}_{~~\mu\nu}=(\rho+p_t)V_\mu
V_\nu-p_tg_{\mu\nu}+(p_r-p_t)\eta_\mu\eta_\nu,
\end{eqnarray}
where $V^\mu V_\mu=-\eta^\mu \eta_\mu=1$ and $\eta^\mu V_\mu=0$,
$V_\mu$ is the 4-velocity and $\eta_\mu$ shows the radial spacelike
4-vector which is orthogonal to $V_\mu$. We can express the
energy-momentum tensor as $\mathcal{T}^{(m)}_{~~\mu\nu}=diag(\rho,
-p_r,-p_t, -p_t)$. Using all the above values in field equations
(\ref{1a}), we obtain
\begin{eqnarray}\nonumber
\rho&=&F(T,T_G)-TF_T(T,T_G)+\frac{2\psi'(r)}{r^2}F_T(T,T_G)
-T_GF_{T_G}(T,T_G)-\frac{4}{r}(1\\\nonumber&-&
\frac{\psi(r)}{r})F_{T}'(T,T_G)+\frac{4}{r^3}\left(\frac{5\psi(r)}{r}
-2-\frac{3\psi^2(r)}{r^2}-3\psi'(r)\left(1-
\frac{\psi(r)}{r}\right)\right)\\\label{7}&\times&F_{T_G}'(T,T_G)
+\frac{8}{r^2}\left(1-\frac{\psi(r)}{r}\left(2-\frac{\psi(r)}{r}\right)\right)
F_{T_G}''(T,T_G),\\\nonumber
p_r&=&-F(T,T_G)+\left(T-\frac{2\psi(r)}{r^3}-\frac{4\zeta}{r^3}+\frac{4\psi(r)\zeta}{r^4}\right)F_T(T,T_G)
\\\label{8}&+&T_GF_{T_G}(T,T_G)
+\frac{48\zeta}{r^4}\left(1-\frac{\psi(r)}{r}\right)^2F_{T_G}'(T,T_G),\\\nonumber
p_t&=&-F(T,T_G)+TF_T(T,T_G)+T_GF_{T_G}(T,T_G)+\left(\frac{\psi(r)}{r^3}-\frac{\psi'(r)}{r^2}
-\frac{2\zeta}{r^3}\right.\\\nonumber&+&\left.\frac{\psi(r)\zeta}{r^4}+\frac{\psi'(r)\zeta}{r^3}
+\frac{2\zeta^2}{r^4}-\frac{2\psi(r)\zeta^2}{r^5}
+\frac{4\zeta}{r^3}-\frac{4\psi(r)\zeta}{r^4}\right)F_{T}(T,T_G)\\\nonumber&+&2\left(\frac{1}{r}-\frac{\psi(r)}{r^2}
-\left(1-\frac{\psi(r)}{r}\right)\zeta'\right)F_{T}'(T,T_G)
+\left(\frac{12\psi(r)\zeta}{r^5}-\frac{12\psi^2(r)\zeta}{r^6}\right.\\\nonumber&+&\left.\frac{12\psi'(r)\zeta}{r^4}
+\frac{12\psi(r)\psi'(r)\zeta}{r^5}-\frac{8\zeta^2}{r^5}+\frac{16\psi(r)\zeta^2}{r^6}
-\frac{8\psi^2(r)\zeta^2}{r^7}-\frac{16\zeta}{r^4}\right.\\\nonumber&-&\left.\frac{32\psi(r)\zeta}{r^5}
-\frac{16\psi^2(r)\zeta}{r^6}\right)F_{T_G}'(T,T_G)+\left(\frac{8\zeta}{r^3}
-\frac{16\psi(r)\zeta}{r^4}+\frac{8\psi^2(r)\zeta}{r^5}\right)\\\label{9}&\times&F_{T_G}''(T,T_G),
\end{eqnarray}
where prime shows derivative with respect to $r$. As,
$T_\mathcal{G}$ contains quartic torsion terms and it is of the same
order with $T^2$. Therefore, $T$ and $\sqrt{\beta
T_\mathcal{G}+T^2}$ are of the same order. So, one should use both
in a modified theory. The simplest non-trivial $F(T,T_\mathcal{G})$
model, which does not introduce a new mass scale into the problem
and differs from GR, is the one described as \cite{6a}
\begin{equation}
F(T,T_\mathcal{G})=\alpha\sqrt{\beta
T_\mathcal{G}+T^2}-T,\\\nonumber
\end{equation}
where $\alpha$ and $\beta$ represent dimensionless non-zero coupling
constants. This model represents interesting cosmological behavior
and can reveal the new features of $F(T,T_G)$ gravity. We have taken
the values of $\alpha$ and $\beta$ from \cite{6a} which provides a
detailed analysis of phase space through this model. We choose those
values of the parameters that correspond to dark energy dominated
era. Also, we take those values of the parameter for which the shape
function of Morris and Thorne static spherically symmetric metric
satisfies its properties. We cannot fix or bound the values of
parameters $\alpha$ and $\beta$. For anisotropic case, we give a
detail discussion about the parameters but in the remaining three
cases, the properties of the shape function are only satisfied for
some particular values of the model that are taken from \cite{6a}.
Using this model in Eqs.(\ref{7})-(\ref{9}), we obtain complicated
form of matter energy density and pressure components given in
Appendix \textbf{A}.

\subsection{Energy Conditions}

The idea of energy conditions came from Raychaudhuri equations
together with the condition of attractive gravity \cite{40}. For
timelike $u^\alpha$ and null $k^\alpha$ vector field congruences,
the Raychaudhuri equations are expressed as follows
\begin{eqnarray}\nonumber
\frac{d\Theta}{d\tau}+R_{\alpha\beta}u^\alpha
u^\beta-\omega_{\alpha\beta}\omega^{\alpha\beta}+\sigma_{\alpha\beta}\sigma^{\alpha\beta}
+\frac{1}{3}\Theta^2&=&0,\\\nonumber
\frac{d\Theta}{d\chi}+R_{\alpha\beta}k^\alpha
k^\beta-\omega_{\alpha\beta}\omega^{\alpha\beta}+\sigma_{\alpha\beta}
\sigma^{\alpha\beta}+\frac{1}{2}\Theta^2&=&0,
\end{eqnarray}
where $\omega^{\alpha\beta}$, $\sigma^{\alpha\beta}$ and $\Theta$
represent the vorticity tensor, shear tensor and expansion scalar,
respectively, $\tau$ and $\chi$ are the parameters. The expression
$\Theta<0$ gives the condition of attractive gravity with
$\omega_{\alpha\beta}=0$ which implies that $R_{\alpha\beta}u^\alpha
u^\beta\geq0$ and $R_{\alpha\beta}k^\alpha k^\beta\geq0$. In these
conditions, the effective energy-momentum tensor is substituted in
the place of Ricci tensor, i.e.,
$\mathcal{T}_{\alpha\beta}^{(eff)}u^\alpha u^\beta\geq0$ and
$\mathcal{T}_{\alpha\beta}^{(eff)}k^\alpha k^\beta\geq0$ which shows
the inclusion of effective energy density and effective pressure in
these conditions. The four energy conditions named as NEC, WEC,
dominant (DEC) and strong energy condition (SEC) are given as
\begin{itemize}
\item NEC:\quad $p_n^{(eff)}+\rho^{(eff)}\geq 0$, where $n=1,2,3$.
\item WEC:\quad $p_n^{(eff)}+\rho^{(eff)}\geq0,\quad\rho^{(eff)}\geq0$,
\item DEC:\quad $p_n^{(eff)}\pm\rho^{(eff)}\geq0,\quad\rho^{(eff)}\geq0$,
\item SEC:\quad
$p_n^{(eff)}+\rho^{(eff)}\geq0,\quad\rho^{(eff)}+3p^{(eff)}\geq0$.
\end{itemize}
The violation of NEC for $\mathcal{T}^{(eff)}_{\lambda\mu}$ plays an
important role to keep the wormhole throat open and to make it
traversable. We evaluate the effective NEC only for radial
coordinate from Eqs.(\ref{7}) and (\ref{8}) as
\begin{equation}\nonumber
p_r^{(eff)}+\rho^{(eff)}=
\left(\frac{\psi'r-\psi}{r^3}+\frac{2}{r}\left(1-\frac{\psi}{r}\right)\lambda'\right).
\end{equation}
We deduce a condition that shows the violation of effective NEC from
the above equation as
\begin{equation}\label{9pN}
\left(\frac{\psi'r-\psi}{r^3}+\frac{2}{r}\left(1-\frac{\psi}{r}\right)\lambda'\right)<0.
\end{equation}
Equation (\ref{9pN}) is the necessary condition for the
traversability of the wormhole. Thus, the violation of NEC for
$\mathcal{T}_{\alpha\beta}^{(eff)}$ gives a possibility for ordinary
matter to fulfil the energy conditions. Hence, the physically
acceptable wormhole solutions can be established in this scenario.

\section{Wormhole Solutions for Various Matter Contents}

In this section, we analyze possible solutions by taking four
different types of fluid. We examine the validity of traversability
condition and also investigate whether
$\mathcal{T}_{\alpha\beta}^{m}$ satisfies the energy bounds or not.

\subsection{Anisotropic Fluid}

First, we discuss the anisotropic fluid model by assuming the
specific form of $\chi(r)$ of the metric function as \cite{27}
\begin{equation}\nonumber
\chi(r)=-\ln\left(1-\left(\frac{r_0}{r}\right)^{m+1}\right),
\end{equation}
where $r_0$ and $m$ are arbitrary constants. Since
$e^{\chi(r)}=\left(1-\frac{\psi(r)}{r}\right)^{-1}$, so the shape
function becomes
\begin{equation}\label{13}
\psi(r)=\frac{r_{0}^{m+1}}{r^m}.
\end{equation}
\begin{figure}\center
\epsfig{file=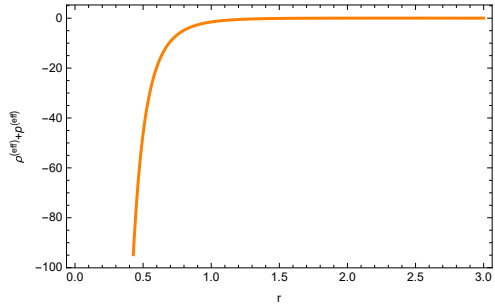,width=0.47\linewidth}\caption{Plot of
$\rho^{(eff)}+p_r^{(eff)}$ versus $r$ for
$\psi(r)=\frac{r_0^{\frac{3}{2}}}{\sqrt{r}}.$}
\end{figure}
It is easy to check that $\psi(r)$ meets all the conditions
necessary to establish a shape function. It satisfies the flaring
out condition $\psi'(r_0)<1$ for $m>1,~\psi(r_0)=r_0$. The condition
of asymptotically flatness is satisfied for the shape function.
Clearly, $\psi(r)$ is characterized through different values of $m$
and can provide meaningful results discussed in literature. Lobo and
Oliveira \cite{17} assumed the above shape function by choosing
$m=1,-\frac{1}{2}$ and investigated the wormhole solutions in $f(R)$
gravity. Pavlovic and Sossich \cite{29} studied the presence of
wormholes in the absence of exotic matter by taking $m=\frac{1}{2}$
in $f(R)$ gravity.

We take the appropriate parameters involved in the above equations
to check the validity of NEC and WEC. We consider different shape
functions for $m=1/2,1,-3$.
\begin{itemize}
\item $\psi(r)=\frac{r_0^{\frac{3}{2}}}{\sqrt{r}}$
\end{itemize}
\begin{figure}\center
\epsfig{file=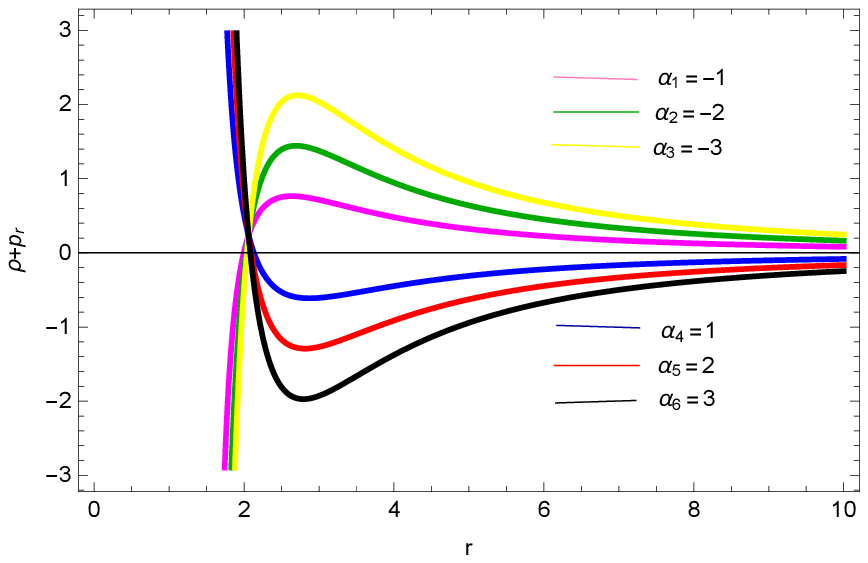,width=0.47\linewidth}
\epsfig{file=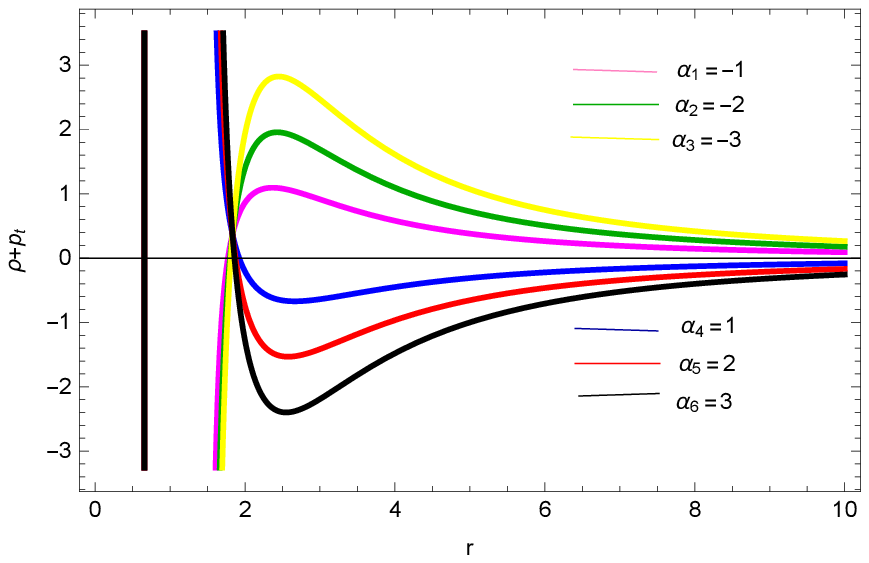,width=0.47\linewidth}
\epsfig{file=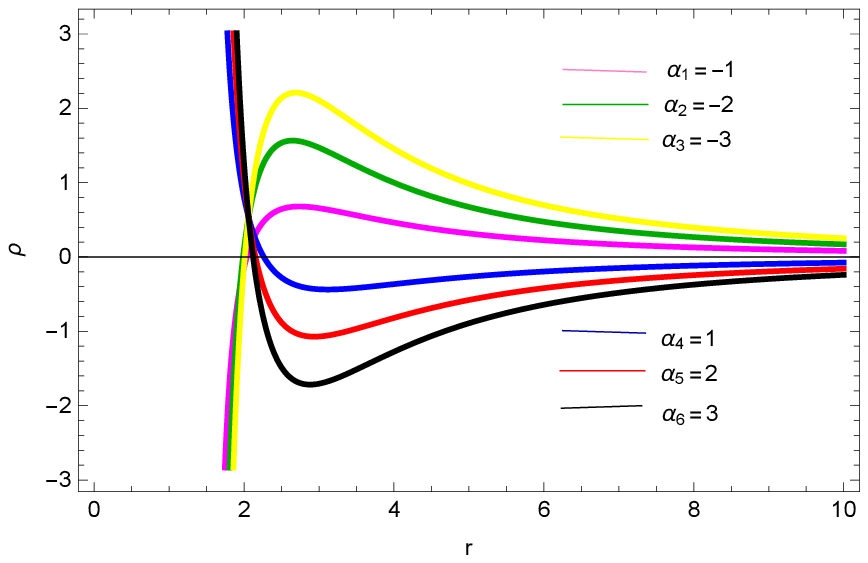,width=0.47\linewidth}\caption{Plots of
$\rho+p_r$, $\rho+p_t$ and $\rho$ versus $r$ for $\beta=1$.}
\end{figure}
\begin{figure}\center
\epsfig{file=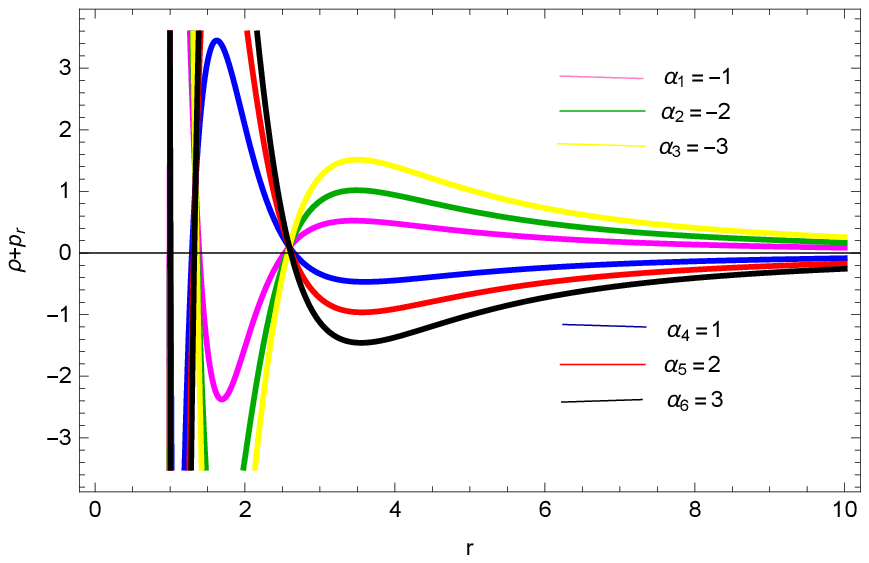,width=0.47\linewidth}
\epsfig{file=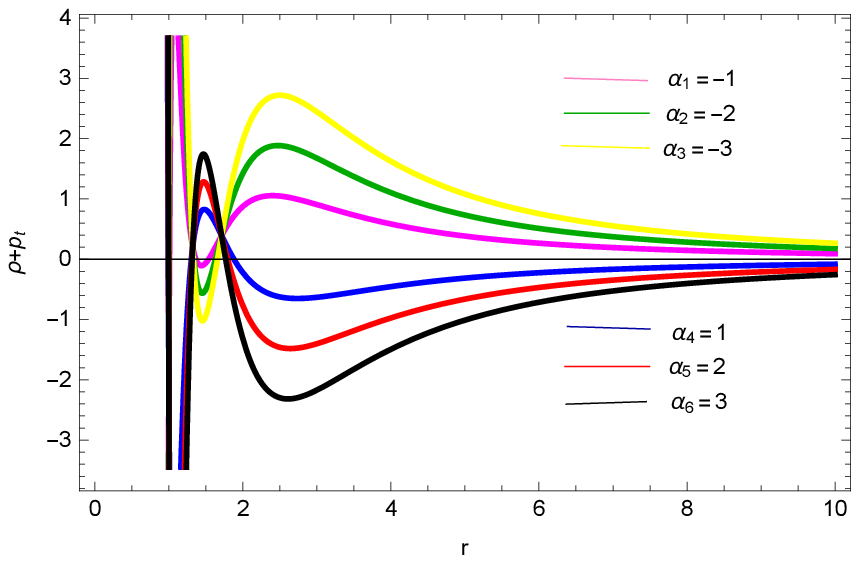,width=0.47\linewidth}
\epsfig{file=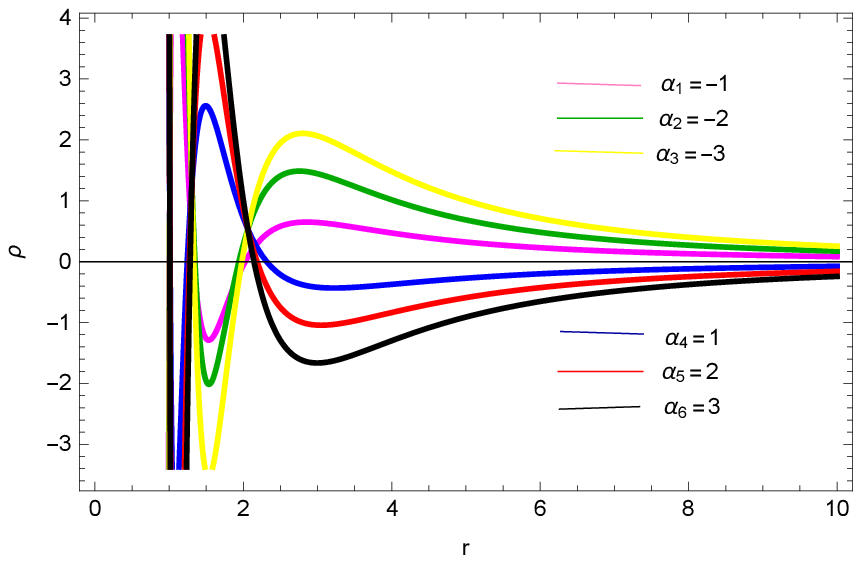,width=0.47\linewidth}\caption{Plots of
$\rho+p_r$, $\rho+p_t$ and $\rho$ versus $r$ for $\beta=-1$.}
\end{figure}
We discuss effective NEC by substituting the above shape function in
Eq.(\ref{9pN}) by taking $m=\frac{1}{2}$,~$r_0=1$ and $\zeta=1$. In
Figure \textbf{1}, the graphical behavior shows the violation of
effective NEC. There appears a possibility for ordinary matter to
satisfy the NEC. So, we investigate NEC and WEC for ordinary matter
by substituting the above shape function in
Eqs.(\ref{10})-(\ref{12}). For this purpose, we consider two choices
of coupling constant, $\beta=1,~-1$ with six different values of
$\alpha$. The behavior of $\rho+p_r$,~$\rho+p_t$ and $\rho$ is shown
in Figure \textbf{2}. For $\beta=1$, NEC and
WEC for ordinary matter are satisfied in the following two regions.\\
(i) When $1.6\leq r\leq2.1,~\alpha_4=1,~\alpha_5=2$ and
$\alpha_6=3$, both energy conditions are satisfied for all
$\alpha>1$.\\
(ii) When $r>2,~\alpha_1=-1,~\alpha_2=-2$ and $\alpha_3=-3$, both
are valid for all $\alpha<-1$. In Figure \textbf{2}, it can be
easily seen that $r$ takes the values in the interval $1.6\leq
r\leq10$ for all the six values of $\alpha$. Similar results are
obtained for all $\beta>1$.

The validity regions of NEC and WEC for $\beta=-1$ are shown in
Figure \textbf{3}. The positive behavior of $\rho+p_r$ is
obtained in the following three regions.\\
(i) When $1.2<r<1.4$ and $r>2.7$ for $\alpha_1=-1,~\alpha_2=-2$ and
$\alpha_3=-3$.\\
(ii) When $1.3<r<2.7$ for $\alpha_4=1,~\alpha_5=2$ and
$\alpha_6=3$.\\
(iii) For $r=1,~\alpha_6=3$.\\
The positive values of $\rho+p_t$ can be obtained for the following
three ranges of the parameters.\\
(i) $1.1<r<1.4$ and $r>1.9$ when $\alpha_1=-1,~\alpha_2=-2$ and
$\alpha_3=-3$.\\
(ii) $1.4<r<2$ when $\alpha_4=1,~\alpha_5=2$ and
$\alpha_6=3$.\\
(iii) For $r=1,~\alpha_6=3$.\\
Finally, $\rho>0$ is satisfied for the following three cases.\\
(i) $1.1<r<1.4$ and $r>1.9$ for $\alpha_1=-1,~\alpha_2=-2$ and
$\alpha_3=-3$.\\
(ii) $1.4<r<2.2$ for $\alpha_4=1,~\alpha_5=2$ and $\alpha_6=3$.\\
(iii) For $r=1$, $\alpha_6=3$.\\
Thus, the following validity regions
for both energy conditions (NEC and WEC) are\\
(i) $1.3<r<1.4$ and $r>2$ when $\alpha_1=-1,~
\alpha_2=-2$, $\alpha_3=-3$.\\
(ii) $1.4<r<2.1$ when $\alpha_4=1,~ \alpha_5=2$ and
$\alpha_6=3$.\\
(iii) For $r=1$, $\alpha_6=3$.\\
\begin{figure}\center
\epsfig{file=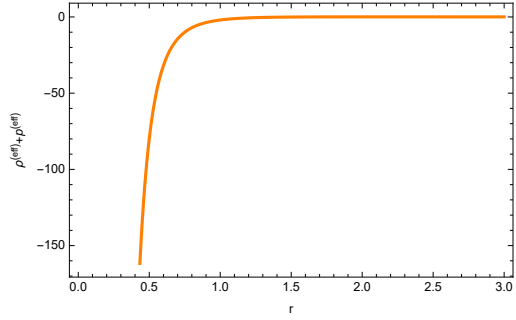,width=0.49\linewidth} \caption{Plot of
$\rho^{(eff)}+p_r^{(eff)}$ versus $r$ for
$\psi(r)=\frac{r_0^2}{r}.$}
\end{figure}
\begin{figure}\center
\epsfig{file=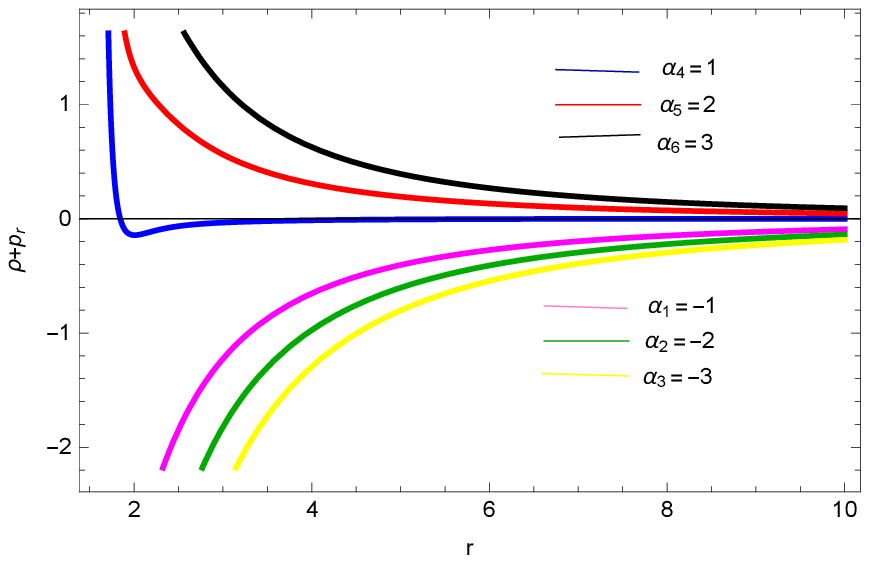,width=0.47\linewidth}
\epsfig{file=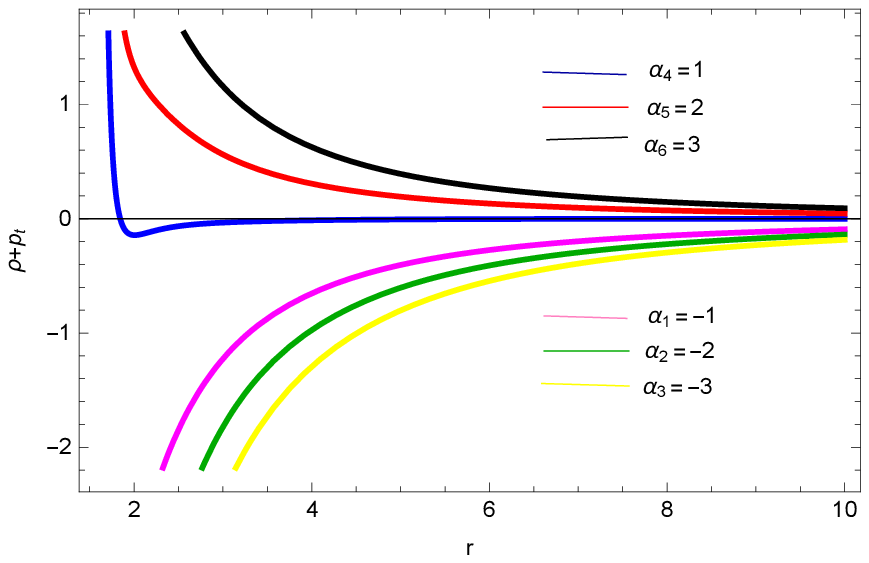,width=0.47\linewidth}
\epsfig{file=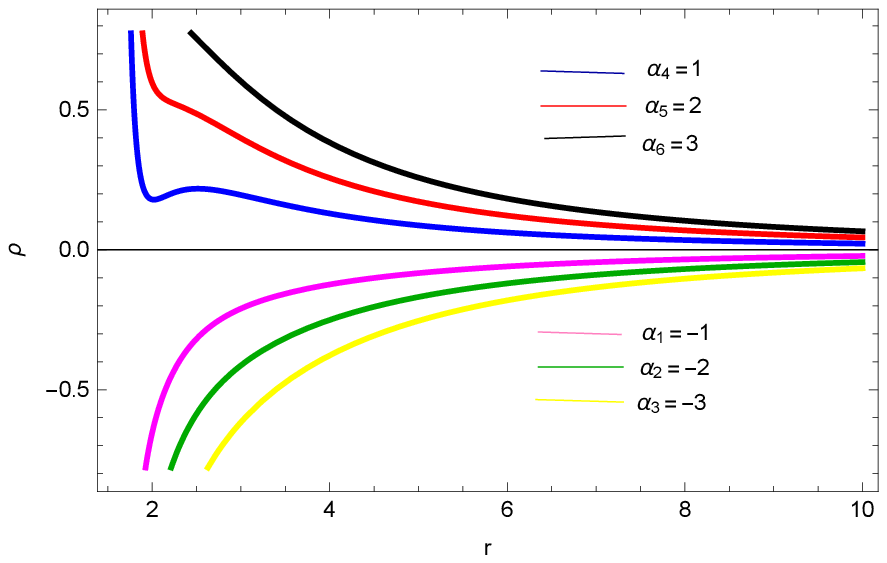,width=0.47\linewidth}\caption{Plots of
$\rho+p_r$, $\rho+p_t$ and $\rho$ versus $r$ for $\beta=1$.}
\end{figure}
\begin{figure}\center
\epsfig{file=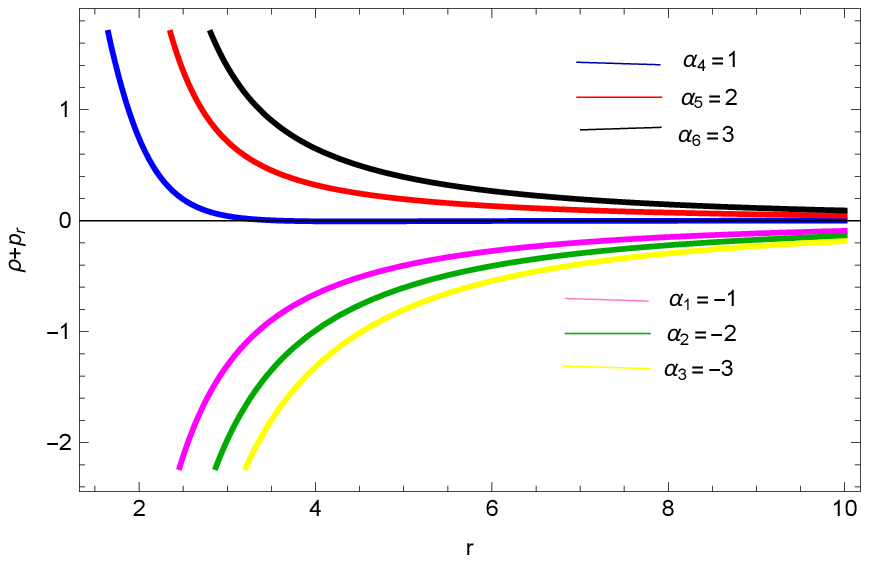,width=0.47\linewidth}
\epsfig{file=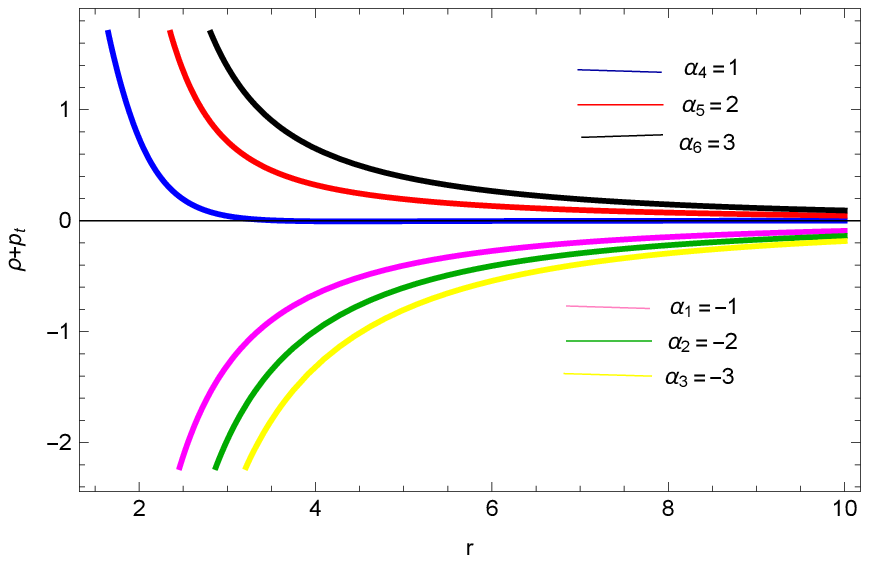,width=0.47\linewidth}
\epsfig{file=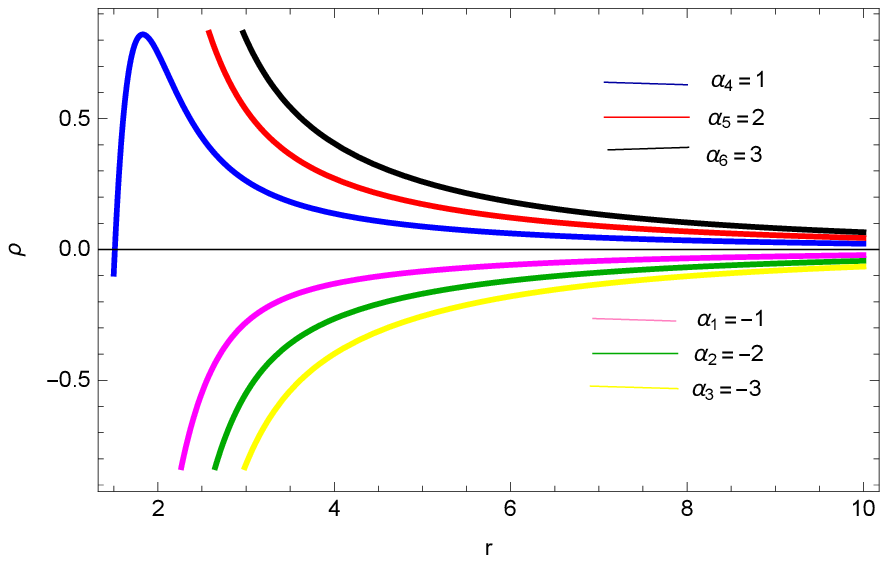,width=0.47\linewidth}\caption{Plots of
$\rho+p_r$, $\rho+p_t$ and $\rho$ versus $r$ for $\beta=-1$.}
\end{figure}

The validity of NEC and WEC for ordinary leads to similar results
for all $\beta<-1,~\alpha<-1,~\alpha>1$. Thus, there exist
physically acceptable wormholes in the above mentioned regions.

\begin{itemize}
\item $\psi(r)=\frac{r_0^2}{r}.$
\end{itemize}
\begin{figure}\center
\epsfig{file=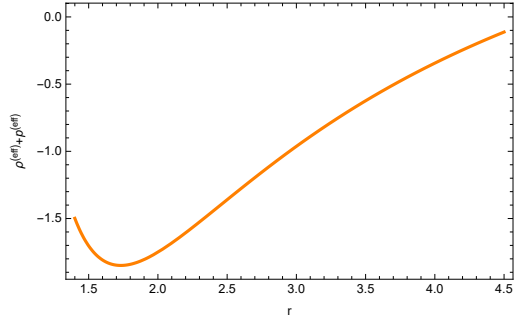,width=0.49\linewidth} \caption{Plot of
$\rho^{(eff)}+p_r^{(eff)}$ versus $r$ for
$\psi(r)=\frac{r^3}{r_0^2}.$}
\end{figure}
For $m=1$, we obtain the above shape function. We check the
condition of traversability given in (\ref{9pN}). In Figure
\textbf{4}, $\rho^{(eff)}+p_r^{(eff)}$ shows negative behavior
versus $r$ that confirms the validity of condition (\ref{9pN}). We
also investigate the NEC as well as WEC for ordinary matter. For
this purpose, we use the above value of shape function in
Eqs.(\ref{10})-(\ref{12}) and find the corresponding results for
$\rho+p_r,~\rho+p_t$ and $\rho$. We discuss these results by
choosing the parametric values $\zeta=1$ and $r_0=1$. Figure
\textbf{5} shows that both $\rho+p_r$ and $\rho+p_t$ represent
positive behavior in the intervals $1.9< r<10$ for $\alpha_5=2$ and
$2.5< r<10$ for $\alpha_6=3$. Also, $\rho+p_r$ and $\rho+p_t$ show
negatively increasing behavior for negative values of $\alpha$. The
energy density indicates positively decreasing behavior for
$\alpha_4=1,~\alpha_5=2,~\alpha_6=3$ and negatively decreasing
behavior for $\alpha_1=-1,~\alpha_2=-2,~\alpha_3=-3$. Thus for
$\beta=1$, NEC and WEC hold in the following regions: (i) $1.9<
r<10,~2.5<r<10$ when $\alpha_5=2$ and $\alpha_6=3$. (ii) $1.6<r<1.8$
for $\alpha_4=1$. When $\beta=-1$, the graphical results of
$\rho+p_r,~\rho+p_t$ and $\rho$ are shown in Figure \textbf{6}. We
can observe that for all $\alpha>1$, the plots of
$\rho+p_r,~\rho+p_t$ and $\rho$ satisfy NEC and WEC. For all
$\alpha<-1$, the behavior of $\rho+p_r,~\rho+p_t$ and $\rho$ do not
meet the energy conditions.
\begin{figure}\center
\epsfig{file=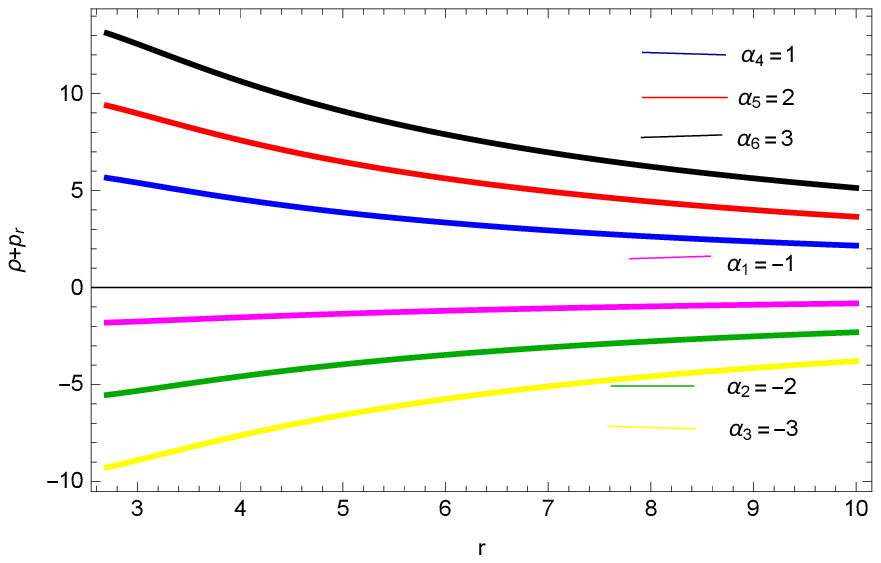,width=0.47\linewidth}
\epsfig{file=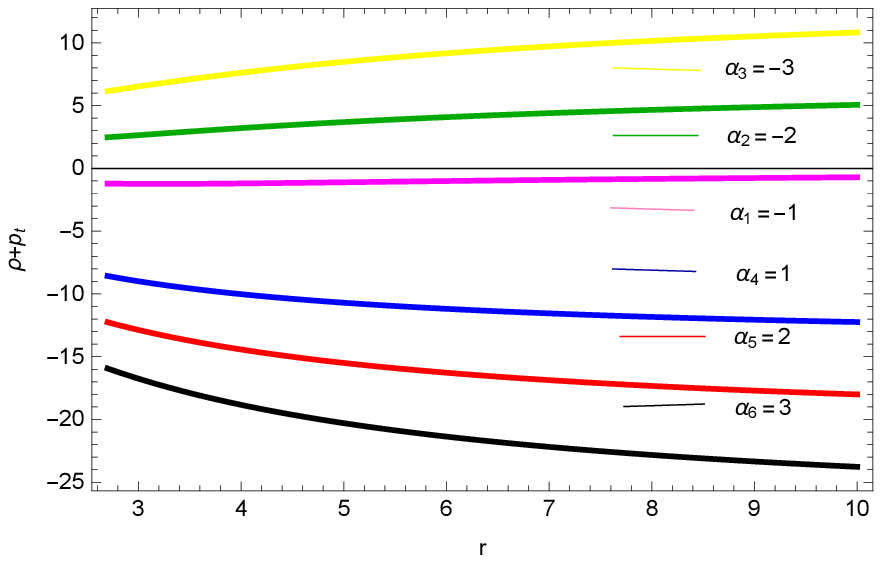,width=0.47\linewidth}
\epsfig{file=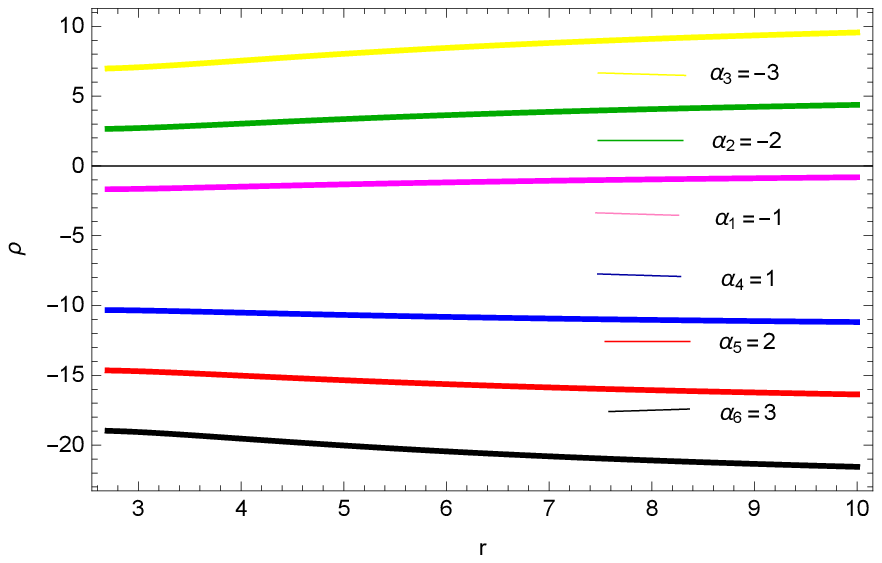,width=0.47\linewidth}\caption{Plots of
$\rho+p_r$, $\rho+p_t$ and $\rho$ versus $r$ for $\beta=1$.}
\end{figure}
\begin{figure}\center
\epsfig{file=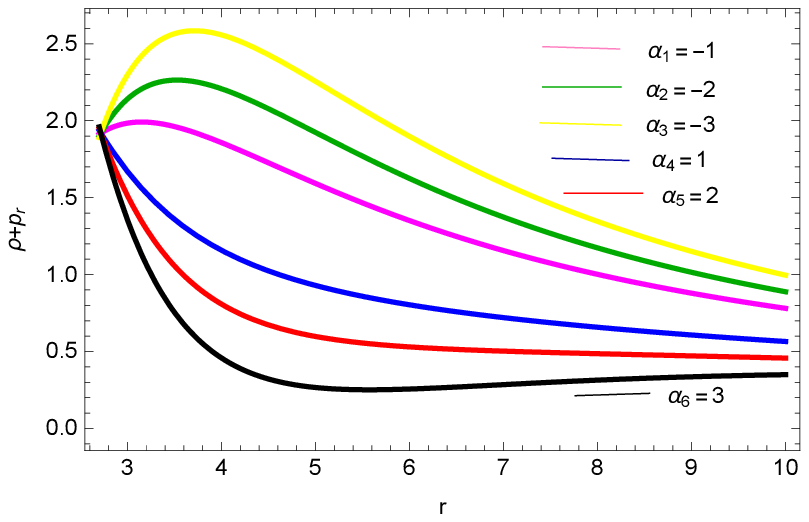,width=0.47\linewidth}
\epsfig{file=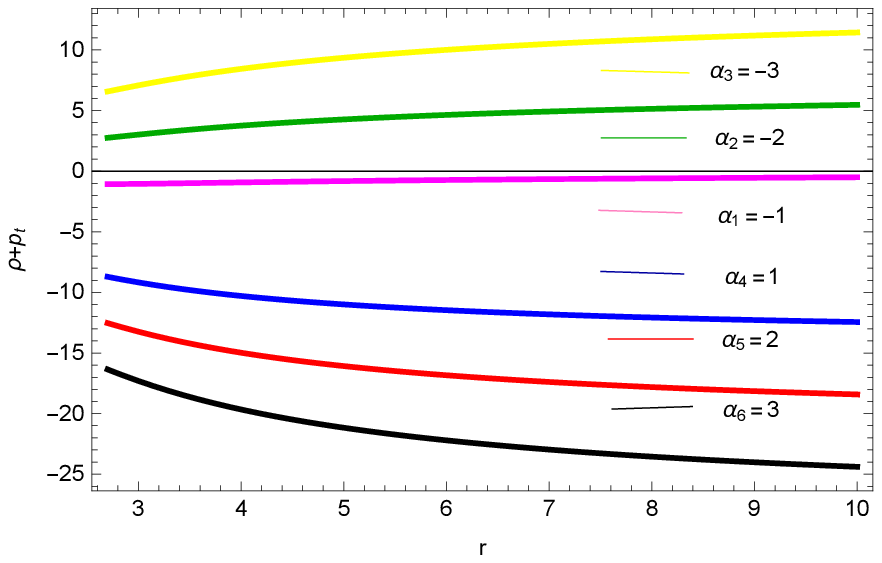,width=0.47\linewidth}
\epsfig{file=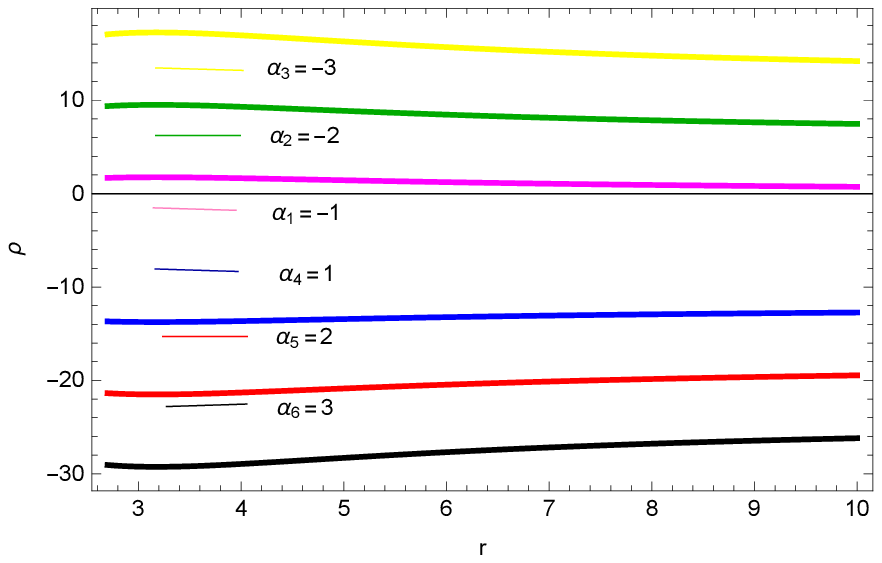,width=0.47\linewidth}\caption{Plots of
$\rho+p_r$, $\rho+p_t$ and $\rho$ versus $r$ for $\beta=-1$.}
\end{figure}
\begin{itemize}
\item $\psi(r)=\frac{r^3}{r_0^2}.$
\end{itemize}

Here, we set $m=-3$ to have the above shape function. The graph of
$\rho^{(eff)}+p_r^{(eff)}$ is shown in Figure \textbf{7}. It can be
observed that the effective NEC does not hold for $m=-3$. We check
the behavior of NEC and WEC for ordinary matter. The corresponding
values of $\rho+p_r,~\rho+p_t$ and $\rho$ are shown in Figure
\textbf{8}. We discuss their behavior by choosing different values
of the parameters $\alpha,~\beta$ and $r$. For $\beta=1$, we see
that $\rho+p_r>0$ for $r\geq2.7,~\alpha_4=1,~\alpha_5=2$ and
$\alpha_6=3$, i.e., for all $\alpha>1$. We find that $\rho+p_t>0$
and $\rho>0$ for $r\geq2.7,~\alpha_2=-2$ and $\alpha_3=-3$. We also
check that their values remain positive for all $\alpha>-1.5$. There
is no similar region between $\rho+p_r$ and $\rho+p_t$, hence NEC
and WEC do not hold for ordinary matter. For $\beta=-1$, both energy
conditions are valid if $r>2.7$ and $\alpha>-2$ as shown in Figure
\textbf{9}. Similar results hold for all $\beta<-1$. Thus, there
exists a realistic wormhole for $r>2.7$ and for all $\alpha>-2$ and
$\beta<-1$.

\subsection{Isotropic Fluid}
\begin{figure}\center
\epsfig{file=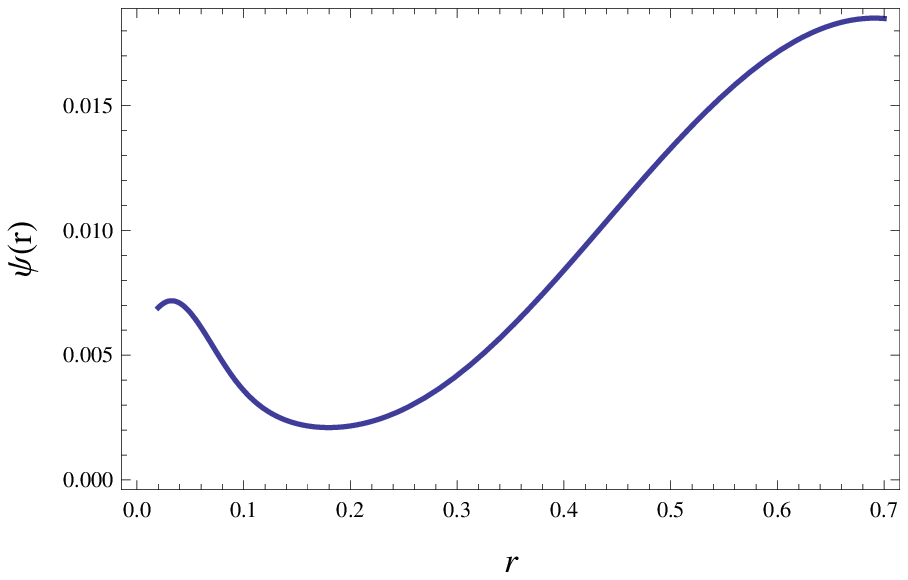,width=0.49\linewidth}
\epsfig{file=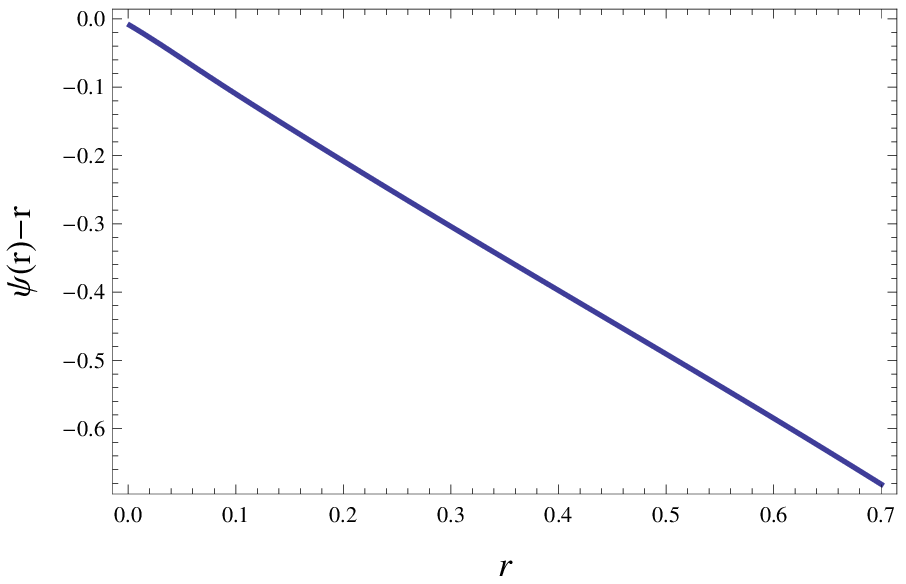,width=0.49\linewidth}
\epsfig{file=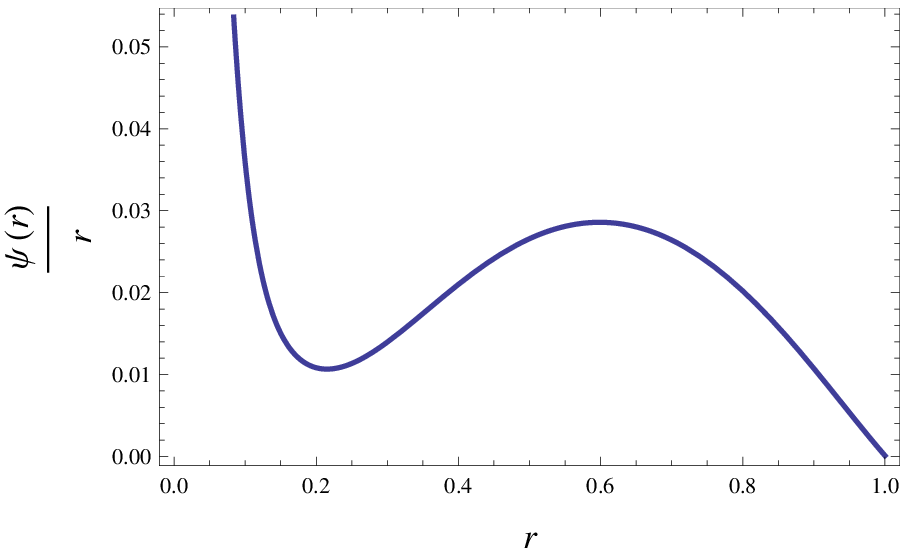,width=0.49\linewidth}
\epsfig{file=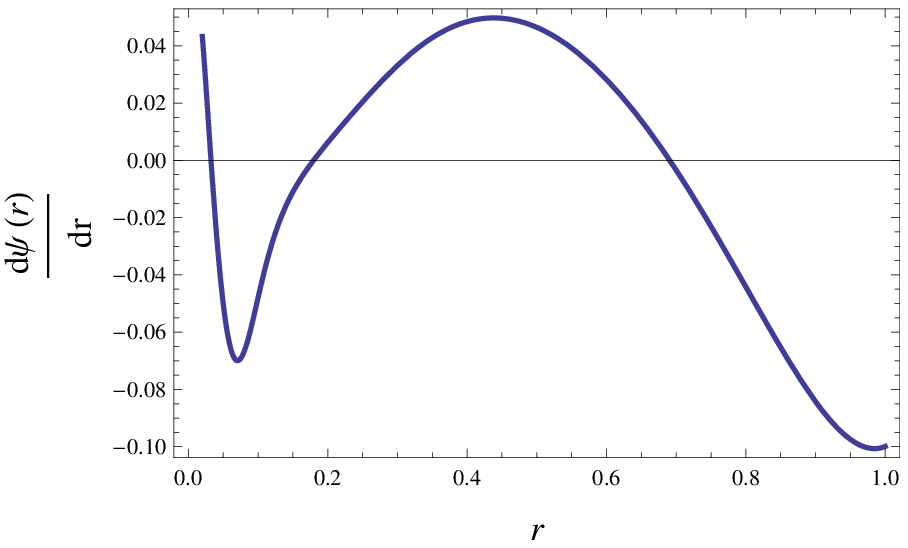,width=0.49\linewidth}\caption{Plots of
$\psi(r)$, $\psi(r)-r$, $\frac{\psi(r)}{r}$ and
$\frac{d\psi(r)}{dr}$ versus $r$ for isotropic case.}
\end{figure}
The equation for isotropic fluid ($p=p_{r}=p_{t}$) is obtained from
Eqs.(\ref{11}) and (\ref{12}) given in Appendix \textbf{A}.

This equation is highly nonlinear which cannot be solved
analytically. We discuss the behavior of shape function and energy
conditions numerically to analyze wormhole solutions for
$\zeta=0.1$, $\alpha=-2$ and $\beta=10$. Figure \textbf{10} shows
that $\psi(r)$ represents the increasing behavior in the interval
$0.2\leq r\leq0.7$ and satisfies the condition $\psi(r)<r$. We find
throat of the wormhole at $r_{th}=0.004$ as $\psi(0.004)=0.00366$.
The asymptotic flatness condition is also satisfied. The plot of the
derivative of shape function indicates that
$\frac{d\psi(r_{rh})}{dr}<1$.
\begin{figure}\center
\epsfig{file=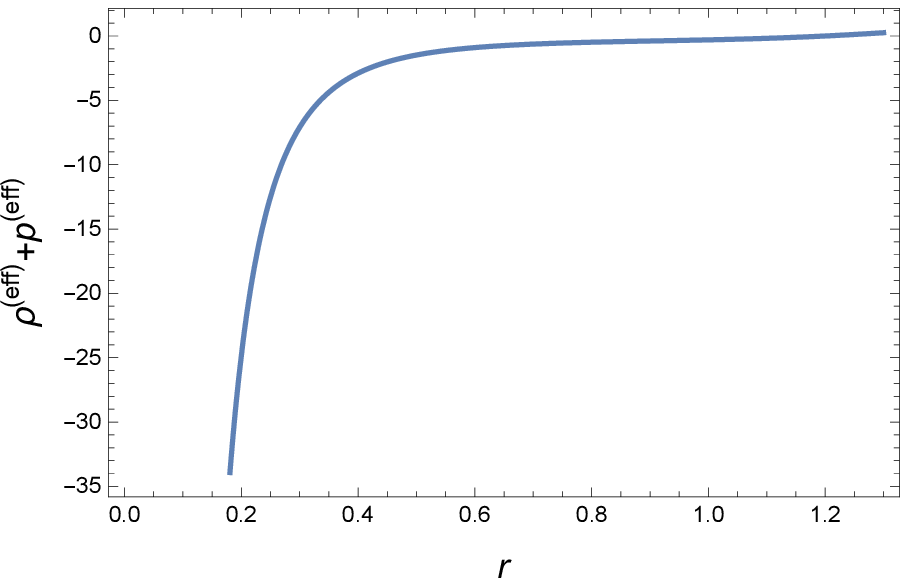,width=0.49\linewidth}
\epsfig{file=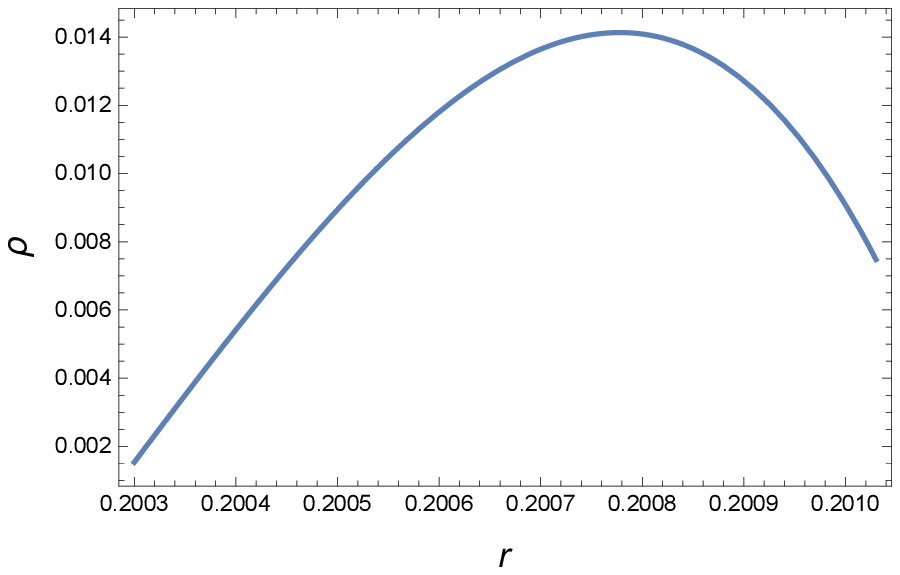,width=0.49\linewidth}
\epsfig{file=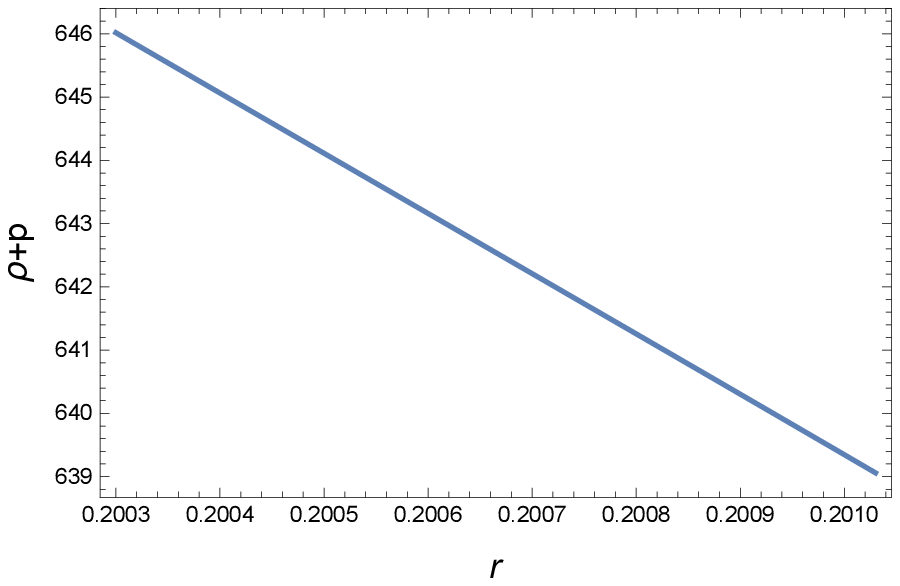,width=0.49\linewidth}\caption{Plots of
$\rho^{(eff)}+p_r^{(eff)}$, $\rho$ and $\rho+p$ versus $r$ for
isotropic case.}
\end{figure}

The upper left panel of Figure \textbf{11} shows the violating
behavior of effective NEC. The upper right panel and lowerpanel of
Figure \textbf{11} represents the evolution of $\rho$ and $\rho+p$
versus $r$. The energy density and for ordinary matter shows
positive behavior in the interval $0.2003\leq r\leq0.2010$. In the
right plot of Figure \textbf{12}, $\rho+p$ lies totally in the
positive region for the same range of $r$. Hence, there may exist a
region of similarity between the two graphs. This shows that both
NEC and WEC satisfy for the isotropic fluid. Hence, there exists a
realistic wormhole in the interval $0.2003\leq r\leq0.2010$ for
$\alpha=-2$ and $\beta=10$.

\subsection{Barotropic Equation of State}

We assume an equation of state which involves energy density and
radial pressure, i.e., $\mu\rho=p_r$, where $\mu$ is the equation of
state parameter. This specific equation of state has been studied in
literature to examine the wormhole solutions \cite{18, 24b, 27}. The
equation for barotropic fluid is obtained from Eqs.(\ref{10}) and
(\ref{11}) given in Appendix \textbf{A}.
\begin{figure}\center
\epsfig{file=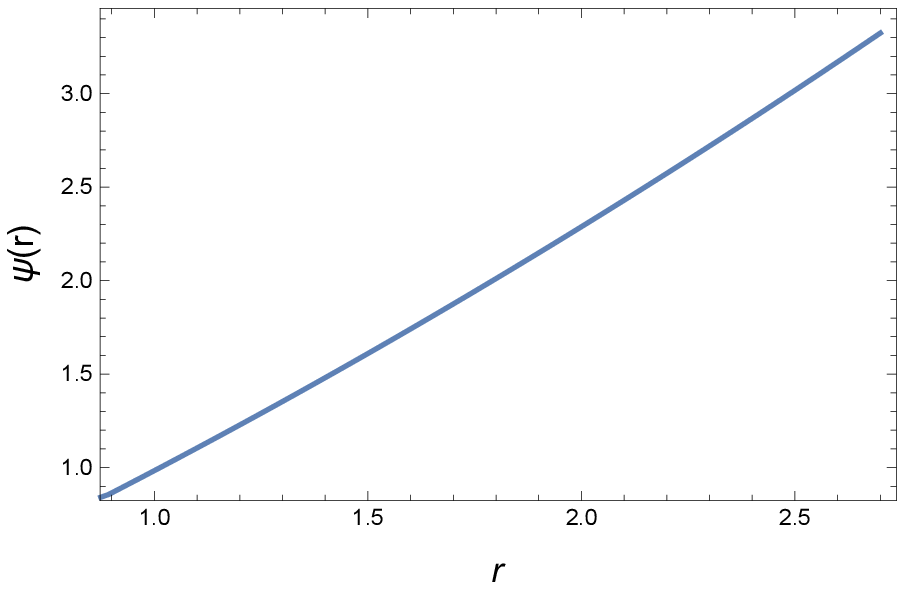,width=0.49\linewidth}
\epsfig{file=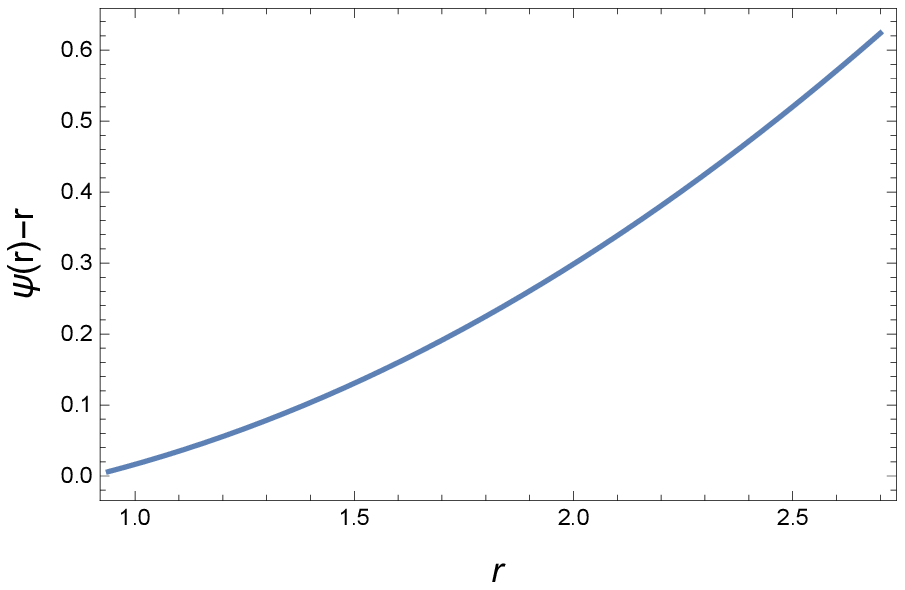,width=0.49\linewidth}
\epsfig{file=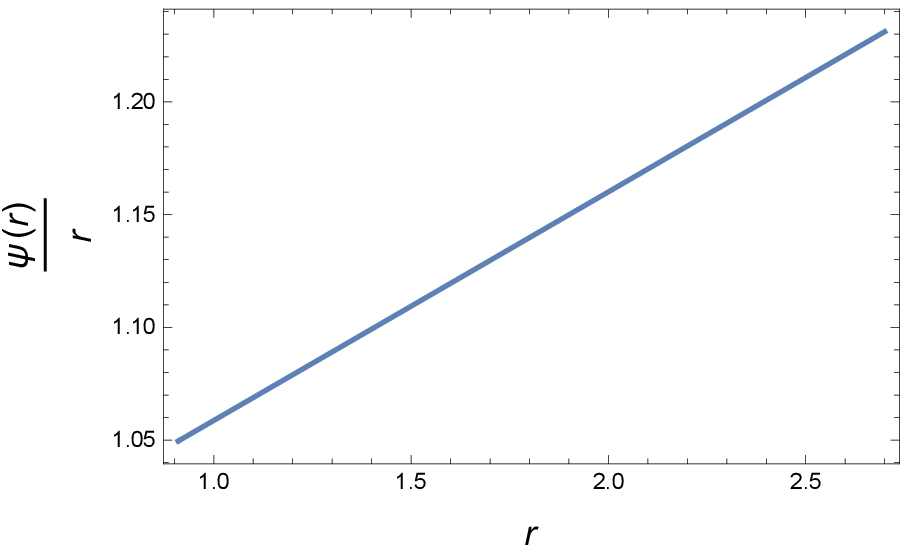,width=0.49\linewidth}\caption{Plots of
$\psi(r)$, $\psi(r)-r$ and $\frac{\psi(r)}{r}$ versus $r$ for
barotropic case.}
\end{figure}
\begin{figure}\center
\epsfig{file=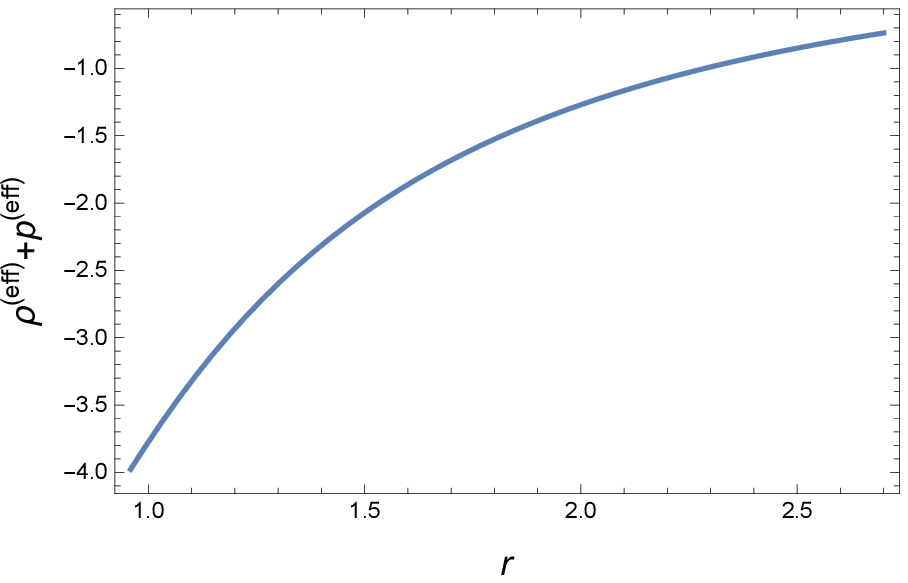,width=0.45\linewidth}
\epsfig{file=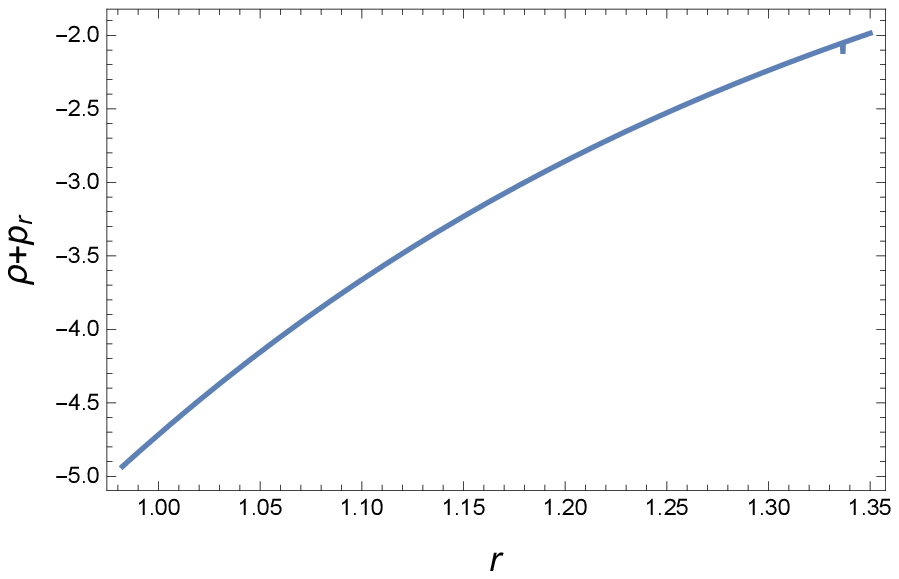,width=0.45\linewidth}
\epsfig{file=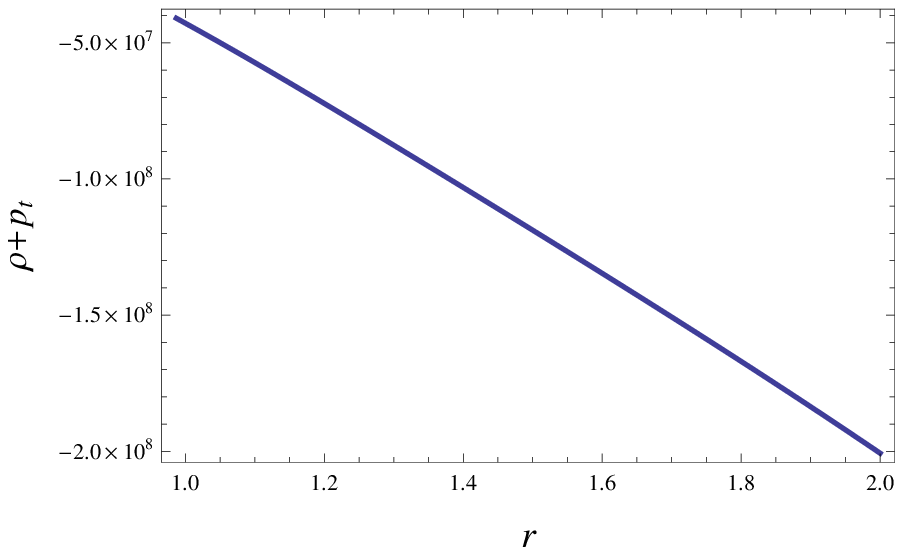,width=0.50\linewidth}
\epsfig{file=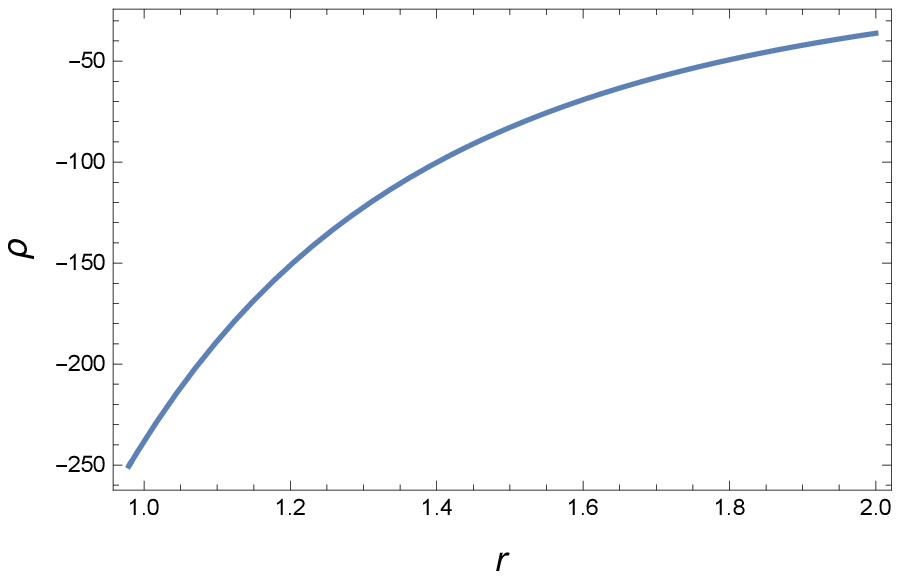,width=0.49\linewidth}\caption{Plots of
$\rho^{(eff)}+p_r^{(eff)}$, $\rho+p_r$, $\rho+p_t$ and $\rho$ versus
$r$ for barotropic case.}
\end{figure}
We solve the above equation numerically to obtain the value of
$\psi(r)$ with $m=\frac{1}{2}$, $\zeta=0.1$, $\alpha=2$, $\mu=0.001$
and $\beta=2$. The left panel of Figure \textbf{12} shows that
$\psi(r)$ increases as the value of $r$ increases. The wormhole
throat is found at very small values of $r$. Also, the plot of
$\frac{\psi(r)}{r}$ shows that the spacetime is not asymptotically
flat. The upper left panel of Figure \textbf{13} represents the
violation of effective NEC. We investigate the behavior of NEC/WEC
for ordinary matter. In Figure \textbf{13}, it can be observed that
$\rho$, $\rho+p_r$ and $\rho+p_t$ exhibit negative values. Thus,
there does not exist realistic wormhole and wormhole geometries are
maintained through exotic matter.

\subsection{Traceless Fluid}
\begin{figure}\center
\epsfig{file=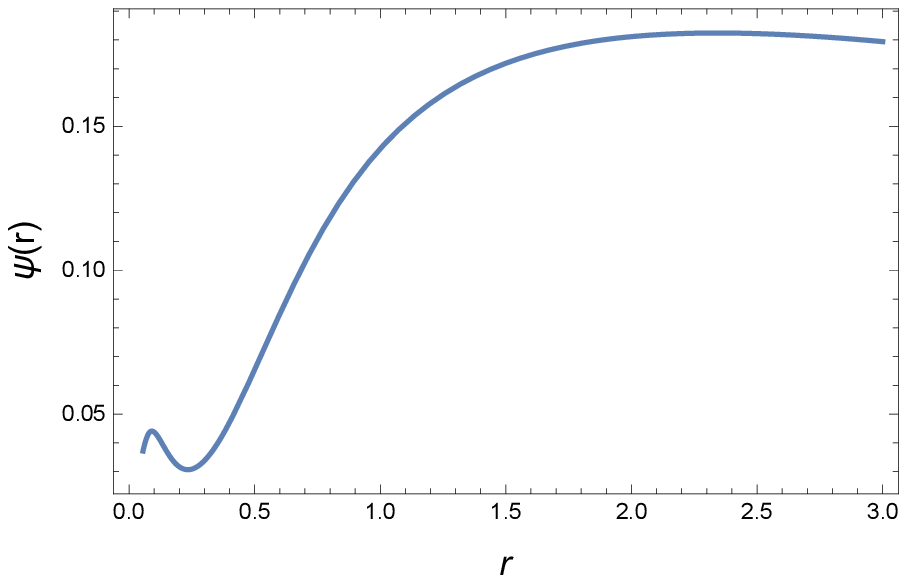,width=0.49\linewidth}
\epsfig{file=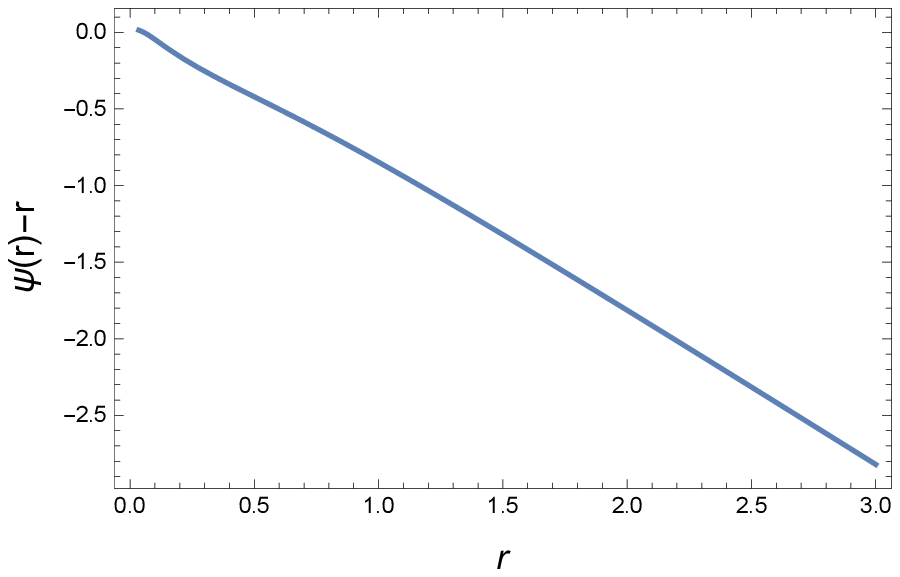,width=0.49\linewidth}
\epsfig{file=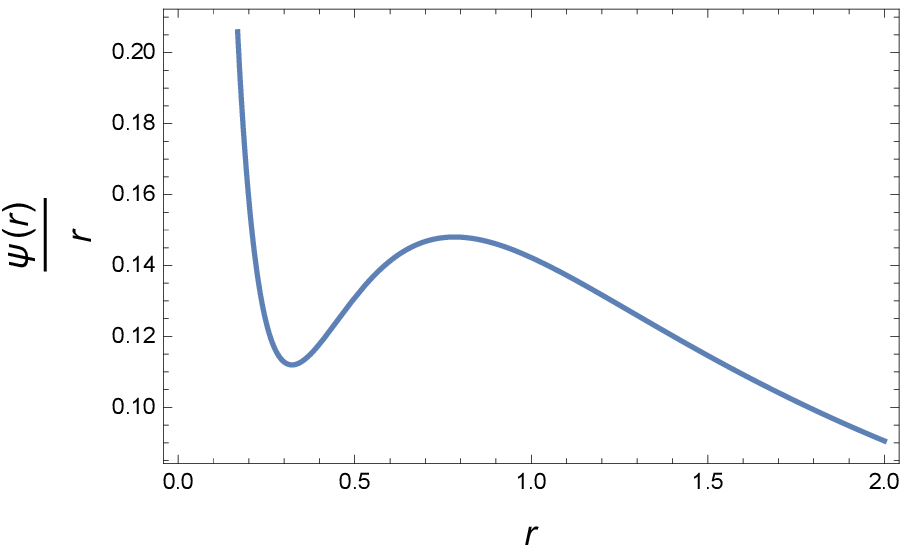,width=0.49\linewidth}
\epsfig{file=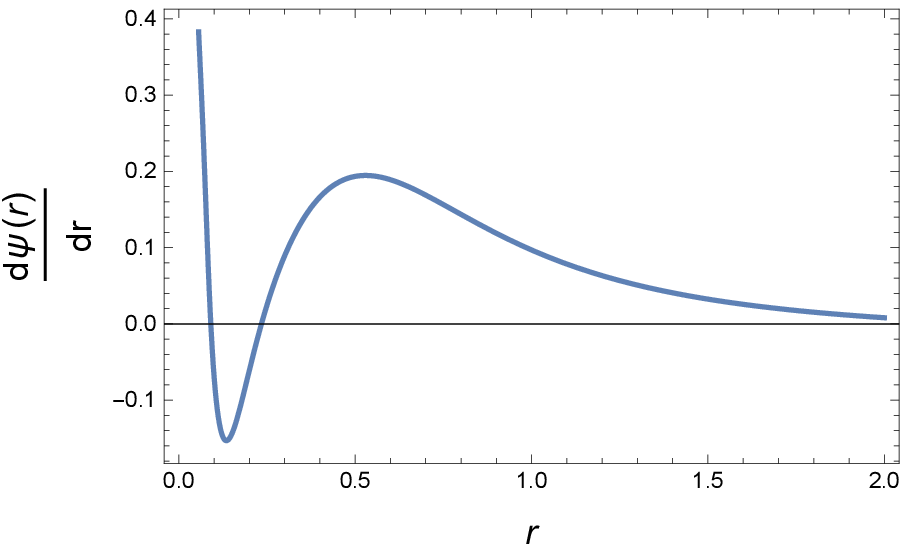,width=0.49\linewidth}\caption{Plots of
$\psi(r)$, $\psi(r)-r$, $\frac{\psi(r)}{r}$ and
$\frac{d\psi(r)}{dr}$ versus $r$ for traceless case.}
\end{figure}
\begin{figure}\center
\epsfig{file=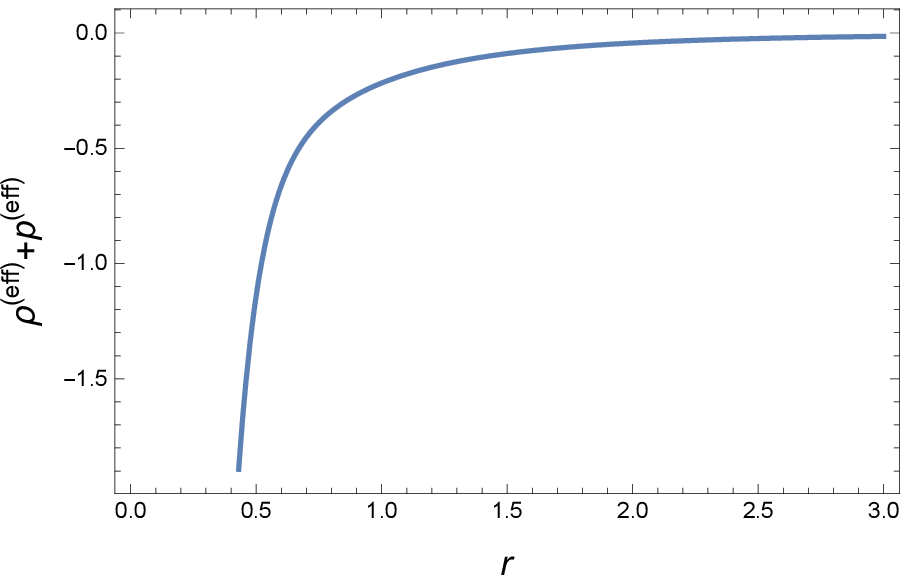,width=0.46\linewidth}
\epsfig{file=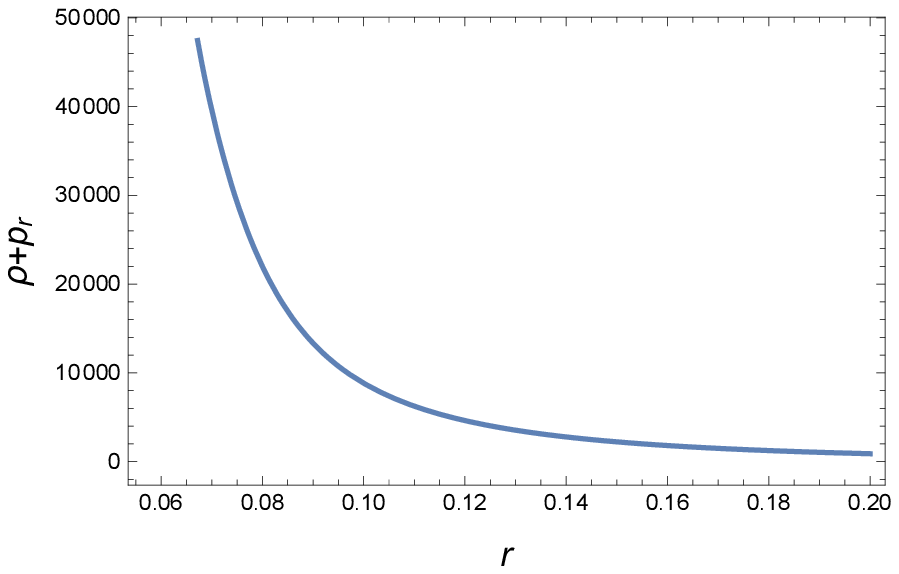,width=0.46\linewidth}
\epsfig{file=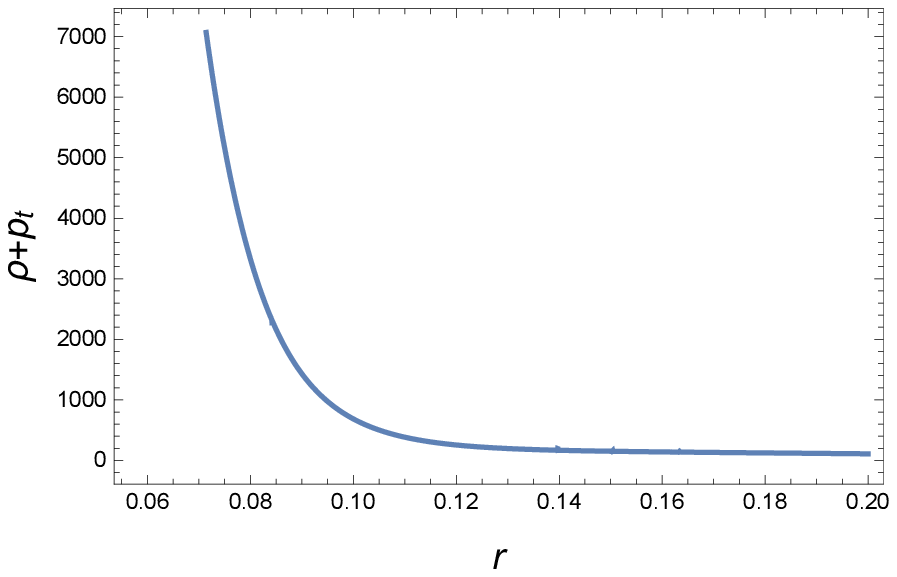,width=0.46\linewidth}
\epsfig{file=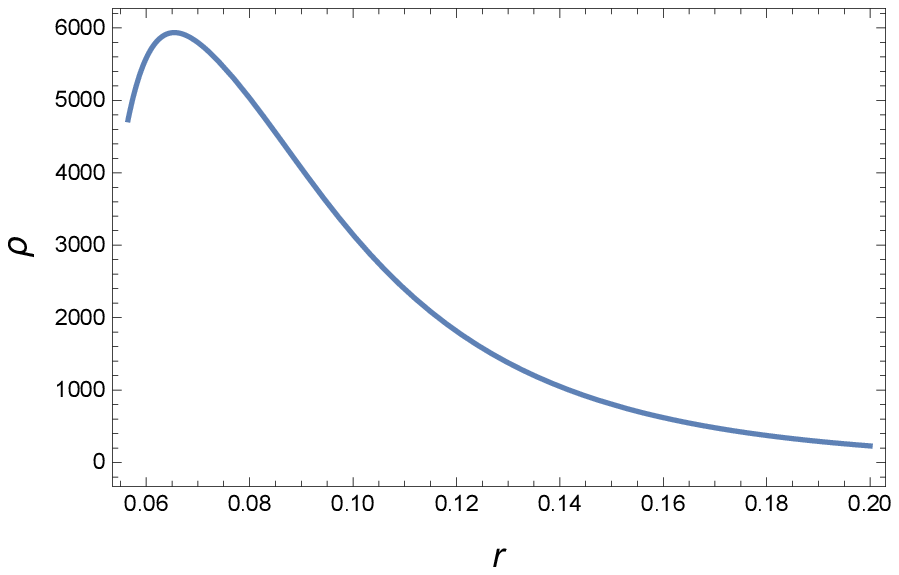,width=0.46\linewidth}\caption{Plots of
$\rho^{(eff)}+p_r^{(eff)}$, $\rho+p_r$, $\rho+p_t$ and $\rho$ versus
$r$ for traceless case.}
\end{figure}
In this case, we investigate the wormhole solutions through an
interesting equation of state for traceless fluid, i.e.,
$2p_t-\rho+p_r=0$ \cite{17, 18}. Therefore,
Eqs.(\ref{10})-(\ref{12}) reduce to the equation given in Appendix
\textbf{A}. Here, we consider the following parameters
$m=\frac{1}{2},~\zeta=0.1,~\alpha=-2$ and $\beta=-2$. Figure
\textbf{14} shows the increasing behavior of $\psi(r)$ and the plot
of $\psi(r)-r$ locates throat of the wormhole at very small values
of $r$. The plot of $\frac{\psi(r)}{r}$ indicates that the spacetime
is asymptotically flat. The derivative of $\psi(r)$ satisfies
$\psi'(r_{th})<1$. The upper left panel of Figure \textbf{15}
indicates the validity of condition (\ref{9pN}). The graphs of
$\rho+p_r,~\rho+p_t$ and $\rho$ (for ordinary matter) express
positive values in the interval $0.06<r<0.2$ as shown in Figure
\textbf{15}. Thus NEC as well as WEC are satisfied for normal matter
in this case. This shows that physically acceptable wormhole exists
in this case for $\alpha$ and $\beta$ is equal to $-2$.

\section{Equilibrium Condition}

Here, we study stability of static wormhole solutions by analyzing
the equilibrium configuration. For this purpose, we deal with the
generalized form of Tolman$-$Oppenheimer$-$Volkov (TOV) equation in
an effective manner given as
\begin{equation}\nonumber
(p^{(eff)}_t-p^{(eff)}_r)\left(\frac{2}{r}\right)-(p^{(eff)}_r
+\rho^{(eff)})\left(\frac{\lambda'(r)}{2}\right)-p'^{(eff)}_r=0.
\end{equation}
We rewrite the above equation as
\begin{equation}\label{14}
-(p^{(eff)}_r+\rho^{(eff)})\left(\frac{M^{(eff)}e^{\frac{\lambda-\chi}{2}}}{r^2}\right)
+(p^{(eff)}_t-p^{(eff)}_r)\left(\frac{2}{r}\right)-p'^{(eff)}_r=0,
\end{equation}
where
$M^{(eff)}=\frac{1}{2}\left(r^2e^{\frac{\chi-\lambda}{2}}\right)\chi'$
is the effective gravitational mass. This equation provides the
equilibrium picture of static wormhole solutions through three
forces namely gravitational force $F_{gf}$, anisotropic force
$F_{af}$ and hydrostatic force $F_{hf}$. The gravitational force
appears as the result of gravitating mass, anisotropic force arises
due to anisotropy of the system and hydrostatic force occurs as a
result of hydrostatic fluid. We can rewrite Eq.(\ref{14}) as
\begin{figure}\center
\epsfig{file=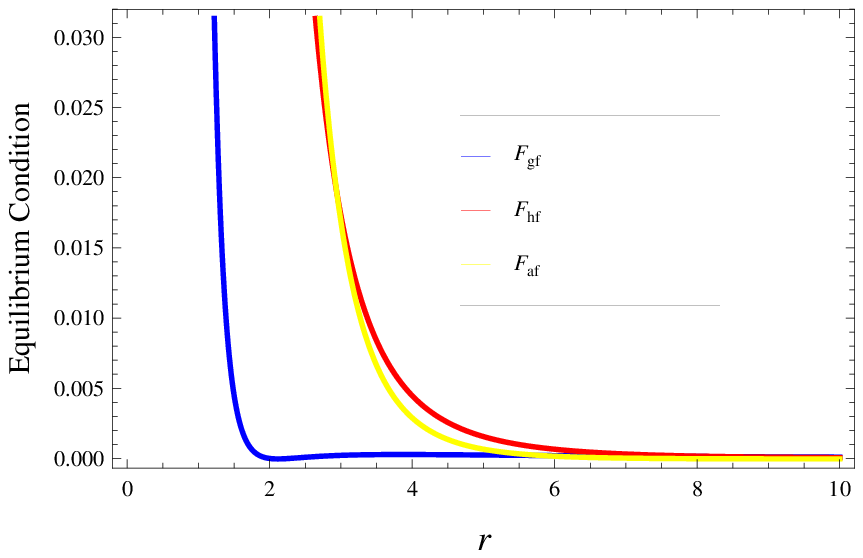,width=0.46\linewidth}
\epsfig{file=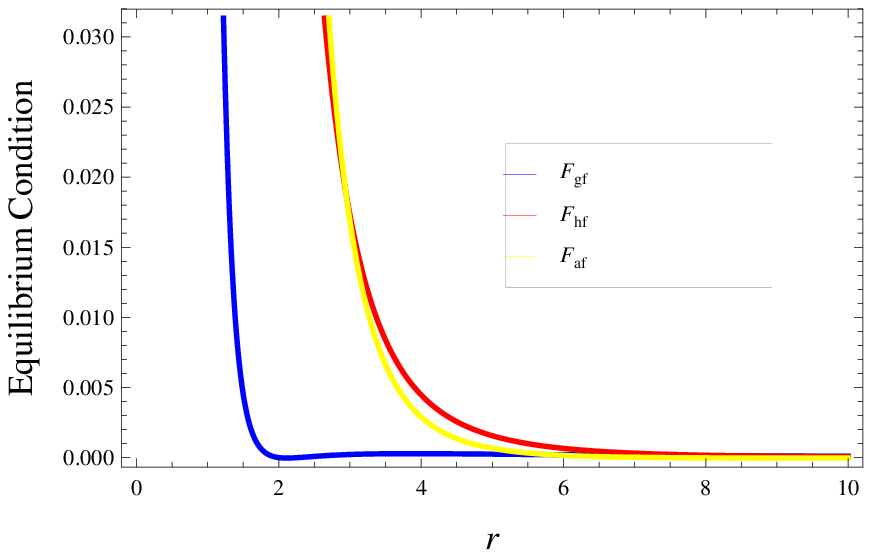,width=0.46\linewidth}\caption{Plot of three
different forces $F_{gf}$, $F_{hf}$ and $F_{af}$ versus $r$ for
$\beta=1$ and $\beta=-1$.}
\end{figure}
\begin{equation}\label{17}
F_{af}+F_{gf}+F_{hf}=0,
\end{equation}
where
\begin{eqnarray}\nonumber
F_{af}&=&(p^{(eff)}_t-p^{(eff)}_r)\left(\frac{2}{r}\right),
F_{gf}=-(p^{(eff)}_r+\rho^{(eff)})\left(\frac{\lambda'(r)}{2}\right),\\\nonumber
F_{hf}&=&-\frac{dp^{(eff)}_r}{dr}.
\end{eqnarray}
Using Eqs.(\ref{10})-(\ref{12}) in (\ref{17}), we obtain an equation
which is solved by applying numerical scheme for $m=\frac{1}{2}$,
$\alpha=-0.0009$ and $\zeta=0.1 $. The graphs of three different
forces are shown in Figure \textbf{16} for $\beta=1$ (left panel)
and $\beta=-1$ (right panel), respectively. In both plots, we see
that the net result of three forces is zero for $r>6$. Hence the
wormhole solutions balance the system implying that these solutions
are stable.

\section{Final Remarks}

In general relativity, the existence of static and traversable
wormholes depends on the presence of exotic matter that violates the
energy conditions. On the other hand, the status may be totally
different in modified theories. In this paper, we have explored the
existence of static and traversable wormhole solutions in the
context of $F(T,T_\mathcal{G})$ gravity for a particular model using
four types of matter contents. For anisotropic case, we have assumed
a shape function to discuss the validity of energy conditions while
for the remaining three cases (isotropic, barotropic and traceless),
we have evaluated the shape function and studied the energy
conditions.

For anisotropic fluid, we have explored energy constraints by using
two values of coupling constant $\beta=1$ and $\beta=-1$. We have
checked the effective NEC for $m=\frac{1}{2},1$ and $-3$. We have
observed that this condition violates for all the three values of
$m$. This violation provides a chance to normal matter to satisfy
the NEC. The energy conditions for normal matter are satisfied for
positive values of $\alpha$ in the specific regions for both
$\beta=1,-1$. We have also checked NEC and WEC for $\beta>1$ and
$\beta<-1$ and obtained similar results. We have discussed three
shape functions with their energy constraints and found results
consistent for $m=-3$ \cite{24c}. For the other three fluids, we
have analyzed solutions numerically to explain the structure of the
shape function and to check the validity of NEC and WEC.

For isotropic and traceless fluids, all the basic requirements are
satisfied by the shape function. The condition for effective NEC is
also satisfied in each case which confirms the presence of
traversable wormhole. The energy conditions for ordinary matter are
also verified for these two fluids. Thus, there exist physically
acceptable wormhole solutions for these fluids. In the barotropic
case, the shape function is not asymptotically flat. The condition
for effective NEC is satisfied but energy conditions for ordinary
matter are violated. Hence, no realistic wormhole solution is found
for barotropic case. We have also explored stability of the wormhole
solutions with anisotropic fluid. It is found that the equilibrium
condition holds and wormhole solutions are stable. We can conclude
that the effective energy-momentum tensor plays a significant role
to violate the energy conditions in $F(T,T_\mathcal{G})$ gravity.
This violation guarantees the presence of traversable wormhole
solutions.

\vspace{0.3cm}

\renewcommand{\theequation}{A\arabic{equation}}
\setcounter{equation}{0}
\section*{Appendix A}

The values of matter energy density and pressure components are
\begin{eqnarray}\nonumber
\rho&=&-Y+\alpha(Y^2+\beta X)^\frac{1}{2}- Y(-1+(\alpha Y)/
(Y^2+\beta
X)^\frac{1}{2})-\frac{1}{2}(X\alpha\beta)/\\\nonumber&\times&
(Y^2+\beta X)^\frac{1}{2}+\frac{1}{r^2}(2\psi'(r)(-1 +(\alpha
Y)/(Y^2+\beta X)^\frac{1}{2}))-\frac{1}{r}
(4(1\\\nonumber&-&\frac{\psi(r)}{r})(-(\alpha Y^2)/(Y^2+\beta Y)
^\frac{3}{2}+\alpha/(Y^2+\beta X)^\frac{1}{2})
\left(-\frac{9}{r^4}\left(1-\frac{\psi(r)}{r}\right)\right.\\\nonumber
&\times&\left.\zeta+\frac{3}{r^3}
\left(-\frac{\psi'(r)}{r}+\frac{\psi(r)}{r^2}\right)\zeta-
\frac{8}{r^5}\left(1-\frac{\psi(r)}{r}\right)\zeta+\frac{2}{r^4}
\left(-\frac{\psi'(r)}{r}\right.\right.\\\nonumber&+&\left.\left.\frac{\psi(r)}{r^2}\right)
\zeta-\frac{4}{r^3}\left(1-\frac{\psi(r)}{r}\right)
+\frac{2}{r^2}\left(-\frac{\psi'(r)}{r}+\frac{\psi(r)}{r^2}\right)\right))-
((5\frac{\psi(r)}{r}-2\\\nonumber&-&3\frac{\psi(r)^2}{r^2}-3\psi'(r)
(1-\frac{\psi(r)}{r}))\alpha\beta^2\left(\frac{12\zeta\psi'^2(r)}{r^6}
+\frac{12\zeta\psi(r)\psi''(r)}{r^6}\right.\\\nonumber&-&\left.
\frac{128\zeta\psi(r)\psi'(r)}{r^7}-\frac{8\zeta^2\psi(r)}{r^7}
+\frac{56\zeta^2\psi(r)}{r^8}+\frac{196\zeta\psi^2(r)}{r^8}+\frac{12\zeta\psi'(r)}{r^7}
\right.\\\nonumber&-&\left.\frac{84\zeta\psi(r)}{r^8}-\frac{8\zeta\psi''(r)}{r^5}
+\frac{40\zeta\psi'(r)}{r^6}
+\frac{16\zeta^2\psi(r)\psi'(r)}{r^8}-\frac{64\zeta^2\psi^2(r)}{r^9}\right))\\\nonumber&\
&(r^3(Y^2+\beta X)\frac{3}{2})
+\frac{1}{r^2}(8(1-\frac{\psi(r)}{r}(2-\frac{\psi(r)}{r}))(\frac{3}{8}
(\alpha\beta^3\left(\frac{12\zeta\psi'^2(r)}{r^6}
\right.\\\nonumber&+&\left.\frac{12\zeta\psi(r)\psi''(r)}{r^6}-
\frac{128\zeta\psi(r)\psi'(r)}{r^7}-\frac{8\zeta^2\psi(r)}{r^7}
+\frac{56\zeta^2\psi(r)}{r^8}+\frac{196\zeta\psi^2(r)}{r^8}\right.\\\nonumber&+&\left.
\frac{12\zeta\psi'(r)}{r^7}
-\frac{84\zeta\psi(r)}{r^8}-\frac{8\zeta\psi''(r)}{r^5}+\frac{40\zeta\psi'(r)}{r^6}
+\frac{16\zeta^2\psi(r)\psi'(r)}{r^8}\right.\\\nonumber&-&\left.\frac{64
\zeta^2\psi^2(r)}{r^9}\right)^2)/(Y^2+\beta
X)
\frac{5}{2}-\frac{1}{4}(\alpha\beta^2(-\frac{240\zeta\psi'(r)}{r^7}
+\frac{12\zeta\psi''(r)}{r^7}\\\nonumber&-&\frac{8\zeta\psi'''(r)}{r^5}
-\frac{8\zeta^2\psi''(r)}{r^7}
-\frac{200\zeta\psi(r)\psi''(r)}{r^7}+\frac{1288\zeta\psi(r)\psi'(r)}{r^8}\\\nonumber
&-&\frac{256\zeta^2\psi(r)\psi'(r)}{r^9}
+\frac{576\zeta^2\psi^2(r)}{r^{10}}+\frac{36\zeta\psi'(r)\psi''(r)}{r^6}
-\frac{200\zeta\psi'(r)}{r^7}
\\\nonumber&+&\frac{12\zeta\psi(r)\psi'''(r)}{r^6}+\frac{112\zeta^2\psi'(r)}{r^8}
-\frac{448\zeta^2\psi(r)}{r^9}
-\frac{1568\zeta\psi^2(r)}{r^9}\\\nonumber&-&\frac{168\zeta\psi'(r)}{r^8}
+\frac{672\zeta\psi(r)}{r^9}
+\frac{80\zeta\psi''(r)}{r^6}+\frac{16\zeta^2\psi'^2(r)}{r^8}+\frac{16\zeta^2\psi(r)
\psi''(r)}{r^8}))\\\label{10}&/&
(Y^2+\beta X^\frac{3}{2})),\\\nonumber p_r&=&Y-\alpha(Y^2+\beta
X)^\frac{1}{2}+(\frac{2}{r^2}(1-\frac{2\psi(r)}{r})+\frac{4}{r^3}(1-
\frac{2\psi(r)}{r})\zeta)
(-1+(\alpha\\\nonumber&\times&Y)/(Y^2+\beta X)^\frac{1}{2})+
\frac{1}{2}((\frac{12\zeta\psi(r)\psi'(r)}{r^6}-\frac{8\zeta^2\psi(r)}{r^7}
-\frac{28\zeta\psi^2(r)}{r^7}\\\nonumber&+&\frac{12\zeta\psi(r)}{r^7}
-\frac{8\zeta\psi'(r)}{r^5}
+\frac{8\zeta^2\psi^2(r)}{r^8})\alpha\beta) /(Y^2+\beta
X)^\frac{1}{2}+(6\zeta(1\\\nonumber&-&\frac{2\psi(r)}{r})^2\alpha\beta^2
(\frac{12\zeta\psi'^2(r)}{r^6}
+\frac{12\zeta\psi'(r)}{r^7}-\frac{12\zeta\psi(r)}{r^8}-\frac{8
\zeta\psi''(r)}{r^5}+\frac{40\zeta\psi'(r)}{r^6}
\\\label{11}
&+&\frac{16\zeta^2\psi'(r)}{r^8}-\frac{64\zeta^2\psi^2(r)}{r^9}))/(r^4(Y^2+\beta
X)^\frac{3}{2}),\\\nonumber p_t&=&Y-\alpha(Y^2+\beta
X)^\frac{1}{2}+(\frac{1}{r}(2-\frac{\psi(r)}{r}
-\frac{\psi'(r)}{r})(\frac{\zeta}{r^2}+\frac{1}{r})+2(1-\frac{\psi(r)}{r})\\\nonumber&\times&
(\frac{\zeta^2}{r^4}-\frac{2\zeta}{r^3}))(-1+(\alpha Y)/(Y^2+\beta
X)^\frac{1}{2})+\frac{1}{2}(X\alpha\beta)/(Y^2+\beta X)^\frac{1}{2}
\\\nonumber&+&2(1-\frac{\psi(r)}{r})(\frac{\zeta}{r^2}+\frac{1}{r})
(-(\alpha Y^2)/(Y^2+\beta X)^\frac{3}{2}+\alpha/(Y^2+\beta
X)^\frac{1}{2})
\\\nonumber&\times&(-\frac{9}{r^4}
(1-\frac{\psi(r)}{r}\zeta+\frac{3}{r^3}(-\frac{\psi'(r)}{r}
+\frac{\psi(r)}{r^2})\zeta-\frac{8}{r^5}
(1-\frac{\psi(r)}{r})\zeta+\frac{2}{r^4}(-\frac{\psi'(r)}{r}
\\\nonumber&+&\frac{\psi(r)}{r^2})\zeta-\frac{4}{r^3}
(1-\frac{\psi(r)}{r})+\frac{2}{r^2}(-\frac{\psi'(r)}{r}
+\frac{\psi(r)}{r^2}))-\frac{1}{4}((\frac{12\zeta}{r^4}(\psi'(r)
+\frac{\psi(r)}{r^2})\\\nonumber&-&\frac{8}{r}(\frac{\zeta^2}{r^4}
-\frac{2\zeta}{r^3})(1-\frac{\psi(r)}{r})^2)\alpha\beta^2
\left(\frac{12\zeta\psi'^2(r)}{r^6}
+\frac{12\zeta\psi(r)\psi''(r)}{r^6}\right.\\\nonumber
&+&\left.\frac{56\zeta^2\psi(r)}{r^8}-\frac{8\zeta^2\psi(r)}{r^7}
-\frac{128\zeta\psi(r)\psi'(r)}{r^7}+\frac{196\xi\psi^2(r)}{r^8}
+\frac{12\zeta\psi'(r)}{r^7}
\right.\\\nonumber&-&\left.\frac{84\zeta\psi(r)}{r^8}
+\frac{8\zeta\psi''(r)}{r^5}+\frac{40\zeta\psi'(r)}{r^6}
+\frac{16\zeta^2\psi(r)\psi'(r)}{r^8}-\frac{64\xi^2\psi^2(r)}{r^9}\right))\\\nonumber
&/&(Y^2+\beta X)^\frac{3}{2}
-\frac{1}{r^3}(8\xi(1-\frac{\psi(r)}{r})^2(\frac{3}{8}
(\alpha\beta^3\left(\frac{12\zeta\psi'^2(r)}{r^6}
+\frac{12\zeta\psi'(r)}{r^7}\right.\\\nonumber&-&\left.
\frac{128\zeta\psi(r)\psi'(r)}{r^7}+\frac{12\zeta\psi(r)\psi''(r)}{r^6}
-\frac{8\zeta^2\psi(r)}{r^7}
+\frac{56\zeta^2\psi(r)}{r^8}\right.\\\nonumber&+&\left.\frac{196\zeta\psi^2(r)}{r^8}
-\frac{84\zeta\psi(r)}{r^8}-\frac{8\zeta\psi''(r)}{r^5}+\frac{40\zeta\psi'(r)}{r^6}
+\frac{16\zeta^2\psi(r)\psi'(r)}{r^8}\right.\\\nonumber&-&\left.
\frac{64\zeta^2\psi^2(r)}{r^9}\right)^2)/(Y^2+\beta X)^\frac{5}{2}
-\frac{1}{4}(\alpha\beta^2Y+\alpha\beta X)
(-\frac{240\zeta\psi'(r)}{r^7}
\\\nonumber&+&\frac{12\xi\psi''(r)}{r^7}-\frac{8\zeta\psi'''(r)}{r^5}
-\frac{8\zeta^2\psi''(r)}{r^7}
-\frac{200\zeta\psi(r)\psi''(r)}{r^7}+\frac{576\zeta^2\psi^2(r)}{r^{10}}
\\\nonumber&-&\frac{256\xi^2\psi(r)\psi'(r)}{r^9}+\frac{1288\zeta\psi(r)\psi'(r)}{r^8}
+\frac{36\zeta\psi'(r)\psi''(r)}{r^6}-\frac{200\zeta\psi'(r)}{r^7}
\\\nonumber&+&\frac{12\zeta\psi(r)\psi'''(r)}{r^6}+\frac{112\zeta^2\psi'(r)}{r^8}
-\frac{448\zeta^2\psi(r)}{r^9}
-\frac{1568\zeta\psi^2(r)}{r^9}\\\nonumber&-&\frac{168\zeta\psi'(r)}{r^8}
+\frac{672\zeta\psi(r)}{r^9}
+\frac{80\zeta\psi''(r)}{r^6}+\frac{16\zeta^2\psi'^2(r)}{r^8}\\\label{12}&+&
\frac{16\zeta^2\psi(r)\psi''(r)}{r^8}))/Y^\frac{3}{2})),
\end{eqnarray}
where
\begin{eqnarray}\nonumber
X&=&\left(\frac{12\zeta
\psi(r)\psi'(r)}{r^6}-\frac{8\zeta^2\psi(r)}{r^7}-\frac{28\zeta
\psi(r)^2}{r^7}+\frac{12\zeta \psi(r)}{r^7}-\frac{8\zeta
\psi'(r)}{r^5}\right.\\\nonumber&+&\left.\frac{8\zeta^2\psi(r)^2}{r^8}\right),\\\nonumber
Y&=&\left(\frac{3\zeta}{r^3}(1-\frac{\psi(r)}{r})+\frac{2\zeta}
{r^4}(1-\frac{\psi(r)}{r})+\frac{2}{r^2}(1-\frac{\psi(r)}{r})\right).
\end{eqnarray}

For isotropic fluid, we obtain the following equation
\begin{eqnarray}\nonumber
&&(-1+(\alpha A)/(A^2+\beta
B)^\frac{1}{2})\left(\frac{2}{r^{2}}\left(1-\frac{2\psi(r)}{r}
\right)+\frac{4\zeta}{r^{3}}\left(1 -\frac{\psi(r)}{r}\right)\right)
+(6\zeta\\\nonumber&\times&\left(1-\frac{\psi(r)}{r}\right)^2
\alpha\beta^2C)/(r^4(A^2 +\beta
B)^\frac{3}{2})-\left(\frac{1}{r}\left(2-\frac{\psi(r)}{r}
-\frac{\psi'(r)}{r}\right)\right.\\\nonumber
&\times&\left. \left(\frac{\zeta}{r^2}
+\frac{1}{r}\right)+2\left(1-\frac{\psi(r)}{r}\right)\left(\frac{\zeta^2}{r^4}
-\frac{2\zeta}{r^3}\right)\right)(-1+(\alpha
A)/(A^2+\beta\\\nonumber&\times&B)^\frac{1}{2})-2\left(1
-\frac{\psi(r)}{r}\right)\left(\frac{\zeta}{r^2}
+\frac{1}{r}\right)(-(\alpha A^2)/(A^2+\beta
B)^\frac{3}{2}+\alpha/(A^2\\\nonumber&+&\beta B)^\frac{1}{2})
\left(-\frac{9\zeta}{r^4}
\left(1-\frac{\psi(r)}{r}\right)+\frac{3\zeta}{r^3}(-\frac{\psi'(r)}{r}
+\frac{\psi(r)}{r^2})-\frac{8\zeta}{r^5}
\left(1-\frac{\psi(r)}{r}\right)\right.\\\nonumber
&+&\left.\frac{2\zeta}{r^4}\left(-\frac{\psi'(r)}{r}+\frac{\psi(r)}{r^2}
\right)-\frac{4}{r^3}
\left(1-\frac{\psi(r)}{r}\right)+\frac{2}{r^2}\left(-\frac{\psi'(r)}{r}
+\frac{\psi(r)}{r^2}\right)\right)\\\nonumber&+&\frac{1}{4}
\left(\left(\frac{12\zeta}{r^4}\left(\psi'(r)-
\frac{\psi(r)}{r^2}\right)\left(1-\frac{\psi(r)}{r}\right)-\frac{8}{r}\left(\frac{\zeta^2}{r^4}
-\frac{2\zeta}{r^3}\right)\left(1\right.\right.\right.\\\nonumber
&-&\left.\left.\left.\frac{\psi(r)}{r}\right)^2\right)
\alpha\beta^2C\right)/(A^2+\beta B)^\frac{3}{2}
+\frac{1}{r^3}(8\zeta\left(1-\frac{\psi(r)}{r}\right)^2(\frac{3}{8}
(\alpha\beta^3C^2)\\\nonumber&/&
(A^2+\beta
B)^\frac{5}{2}-\frac{1}{4}\left(\alpha\beta^2\left(-\frac{240\zeta\psi'(r)}{r^7}
-\frac{8\zeta^2\psi''(r)}{r^7}
-\frac{8\zeta\psi'''(r)}{r^5}
\right.\right.\\\nonumber&+&\left.\left.\frac{12\zeta\psi''(r)}{r^7}
-\frac{200\zeta\psi(r)\psi''(r)}{r^7}
+\frac{1288\zeta\psi(r)\psi'(r)}{r^8}
-\frac{256\zeta^2\psi(r)\psi'(r)}{r^9}\right.\right.\\\nonumber
&+&\left.\left.\frac{36\zeta\psi'(r)\psi''(r)}{r^6}
\frac{200\zeta\psi'^2(r)}{r^7}
+\frac{12\zeta\psi(r)\psi'''(r)}{r^6}+\frac{112\zeta^2\psi'(r)}{r^8}\right.\right.\\\nonumber
&-&\left.\left.
\frac{448\zeta^2\psi(r)}{r^9}\frac{1568\zeta\psi^2(r)}{r^9}-\frac{168\zeta\psi'(r)}{r^8}
+\frac{672\zeta\psi(r)}{r^9}+\frac{80\zeta\psi''(r)}{r^6}
\right.\right.\\\nonumber&+&\left.\left.\frac{16\zeta^2\psi'^2(r)}{r^8}
+\frac{16\zeta^2\psi(r)\psi''(r)}{r^8}
\frac{576\zeta^2\psi^2(r)}{r^{10}}\right)\right)/(A^2+\beta
B)^\frac{3}{2}))=0,
\end{eqnarray}
where
\begin{eqnarray}\nonumber
A&=&\frac{3\zeta}{r^{3}}\left(1-\frac{\psi(r)}{r}\right)+\frac{2\zeta}{r^{4}}
\left(1-\frac{\psi(r)}{r}\right)
+\frac{2}{r^{2}}\left(1-\frac{\psi(r)}{r}\right).\\\nonumber
B&=&\frac{12A\psi(r)\psi'(r)}{r^6}-\frac{8A^2\psi(r)}{r^7}-\frac{28A\psi^2(r)}
{r^7}+\frac{12A\psi(r)}{r^7}-
\frac{8A\psi'(r)}{r^5}\\\nonumber&+&\frac{8A^2\psi^2(r)}{r^8}).\\\nonumber
C&=&\frac{12\zeta(\psi'(r))^2}{r^{6}}+\frac{12\zeta\psi(r)\psi''(r)}{r^{6}}
-\frac{128\zeta\psi(r)\psi'(r)}{r^{7}}
-\frac{8\zeta^2\psi'(r)}{r^{7}}+\frac{56\zeta^2\psi(r)}{r^{8}}\\\nonumber
&+&\frac{196\zeta\psi^2(r)}{r^{8}}+ \frac{12\zeta\psi'(r)}{r^{7}}
-\frac{84\zeta\psi(r)}{r^{8}}-\frac{8\zeta\psi''(r)}{r^{5}}
+\frac{40\zeta\psi'(r)}{r^{6}}-\frac{64\zeta^2\psi^2(r)}{r^{9}}
\\\nonumber&+&\frac{16\zeta^2\psi(r)\psi'(r)\psi}{r^{8}}.
\end{eqnarray}

For barotropic fluid, we obtain the following equation as
\begin{eqnarray}\nonumber
&&A-\alpha (A^{2}+\beta
B^{\frac{1}{2}}+\left(\frac{2}{r^2}\left(1-\frac{2\psi(r)}{r}\right)
+\frac{4\zeta}{r^3}\left(1
-\frac{\psi(r)}{r}\right)\right)(\alpha A\\\nonumber&/&(A^2+\beta
B)^{\frac{1}{2}})+\frac{1}{2}(B\alpha\beta)/ (A^2+\beta
B)^{\frac{1}{2}}+(6\zeta\left(1-\frac{\psi(r)}{r}\right)^2
\alpha\beta^2C)\\\nonumber&/&(r^4(A^2
+\beta B)^\frac{3}{2})-\mu(-A+\alpha(A^2+B)^\frac{1}{2}-A(-1+(\alpha
A)/(A^2+\beta\\\nonumber&\times&B)^\frac{1}{2})
-\frac{1}{2}(B\alpha\beta) /(A^2 +\beta B)^
\frac{1}{2}+\frac{1}{r^2}(2\psi'(r)(-1+(\alpha
A)/(A^2+\beta\\\nonumber&\times&B)^\frac{1}{2}))-\frac{1}{r}(4
\left(1-\frac{\psi(r)}{r}\right)(-(\alpha A^2)/(A^2+\beta
B)^\frac{3}{2}+\alpha/(A^2+\beta\\\nonumber&\times&B)^\frac{1}{2})
\left(-\frac{9\zeta}{r^4}\left(1-\frac{\psi(r)}{r}\right)+\frac{3\zeta}{r^3}
\left(-\frac{\psi'(r)}{r}
+\frac{\psi(r)}{r^2}\right)-\frac{8\zeta}{r^5}\left(1-\frac{\psi(r)}{r}\right)\right.\\\nonumber
&+&\left.
\left.\frac{2\zeta}{r^4}\left(-\frac{\psi'(r)}{r}+\frac{\psi(r)}{r^2}\right)
-\frac{4}{r^3}\left(1-\frac{\psi(r)}{r}\right)
+\frac{2}{r^2}\left(-\frac{\psi'(r)}{r}+\frac{\psi(r)}{r^2}\right)\right)\right)\\\nonumber
&-&\left(\left(\frac{5\psi(r)}{r}-2
-\frac{3\psi^2(r)}{r^2}-3\psi'(r)\left(1-\frac{\psi(r)}{r}\right)\right)\alpha\beta^2C\right)/
(r^3(A^2\\\nonumber&+&\beta B)^\frac{3}{2})+\frac{1}{r^2}(8(1-
\frac{\psi(r)}{r}(2-\frac{\psi'(r)}{r}))(\frac{3}{8}(\alpha\beta^3C^2)/(A^2+_2B)^\frac{5}{2}
-\frac{1}{4}\\\nonumber&\times&(\alpha\beta^2(-\frac{240\zeta\psi'(r)}{r^7}
-\frac{168\zeta\psi'(r)}{r^8}-\frac{8
\zeta^2\psi''(r)}{r^7}
-\frac{8\zeta\psi'''(r)}{r^5}\frac{12\zeta\psi''(r)}{r^7}\\\nonumber
&-&\frac{200\zeta\psi(r)\psi''(r)}{r^7}+\frac{1288
\zeta\psi(r)\psi'(r)}{r^8}
-\frac{256\zeta^2\psi(r)\psi'(r)}{r^9}+\frac{36\zeta\psi'(r)\psi''(r)}{r^6}\\\nonumber
&-&\frac{200\zeta\psi'^2(r)}{r^7}
+\frac{12\zeta\psi(r)\psi'''(r)}{r^6}+\frac{112\zeta^2\psi'(r)}{r^8}
-\frac{448\zeta^2\psi(r)}{r^9}
-\frac{1568\zeta\psi^2(r)}{r^9}\\\nonumber&+&\frac{672\zeta\psi(r)}{r^9}
+\frac{80\zeta\psi''(r)}{r^6}
+\frac{16\zeta^2\psi'^2(r)}{r^8}+\frac{16\zeta^2\psi(r)\psi''(r)}{r^8}
+\frac{576\zeta^2\psi^2(r)}{r^{10}}))\\\nonumber&/&(A^2 +\beta
B)^\frac{3}{2})))=0.
\end{eqnarray}

For traceless fluid, we obtain the following equation as
\begin{eqnarray}\nonumber
&&(6\zeta\left(1-\frac{\psi(r)}{r}\right)^2\alpha\beta^2C)/(r^4(A^2+\alpha_2B)^\frac{3}{2})
+\frac{8}{r^2}
\left(1-\frac{\psi(r)}{r}\right)-\frac{1}{2}\\\nonumber&\times&(B\alpha\beta^2C)
/(A^2+\beta B)^\frac{3}{2}-\frac{1}{r^3}
(16\zeta\left(1-\frac{\psi(r)}{r}\right)^2
(\frac{3}{8}(\alpha\beta^3A^2)/(A^2\\\nonumber&+&\beta
B)^\frac{5}{2}-\frac{1}{4}\left(\alpha\beta^2
\left(-\frac{240\zeta\psi'(r)}{r^7}-\frac{8\zeta^2\psi''(r)}{r^7}
-\frac{8\zeta\psi'''(r)}{r^5}
+\frac{12\zeta\psi''(r)}{r^7}\right.\right.\\\nonumber&-&\left.\left.
\frac{200\zeta\psi(r)\psi''(r)}{r^7}
+\frac{1288\zeta\psi(r)\psi'(r)}{r^8}
-\frac{256\zeta^2\psi(r)\psi'(r)}{r^9}-\frac{200\zeta\psi'^2(r)}{r^7}\right.\right.\\\nonumber
&+&\left.\left. \frac{36\zeta\psi'(r)\psi''(r)}{r^6}
+\frac{12\zeta\psi(r)\psi'''(r)}{r^6}+\frac{112\zeta^2\psi'(r)}{r^8}-
\frac{448\zeta^2\psi(r)}{r^9}\right.\right.\\\nonumber
&-&\left.\left.\frac{1568\zeta\psi^2(r)}{r^9}-\frac{168\zeta\psi'(r)}{r^8}
+\frac{672\zeta\psi(r)}{r^9}+\frac{80\zeta\psi''(r)}{r^6}+
\frac{16\zeta^2\psi'^2(r)}{r^8}\right.\right.\\\nonumber&+&\left.\left.
\frac{16\zeta^2\psi(r)\psi''(r)}{r^8}+
\frac{576\zeta^2\psi^2(r)}{r^10}\right)\right)/(A^2+\beta
B)^\frac{3}{2}))
-\frac{1}{r^2}(8(1-\frac{\psi(r)}{r}\\\nonumber&\times&(2-\frac{\psi(r)}{r}))
(\frac{3}{8}(\alpha\beta^3C^2)/(A^2
+\beta B)^\frac{5}{2}
-\frac{1}{4}(\alpha\beta^2\left(-\frac{240\zeta\psi'(r)}{r^7}\right.\\\nonumber
&-&\left.\frac{8\zeta^2\psi''(r)}{r^7} -\frac{8\zeta\psi'''(r)}{r^5}
+\frac{12\zeta\psi''(r)}{r^7}-\frac{200\zeta\psi(r)\psi''(r)}{r^7}
-\frac{256\zeta^2\psi(r)\psi'(r)}{r^9}\right.\\\nonumber&+&\left.
\frac{1288\zeta\psi(r)\psi'(r)}{r^8}
+\frac{36\zeta\psi'(r)\psi''(r)}{r^6}-\frac{200\zeta\psi'^2(r)}{r^7}
+\frac{12\zeta\psi(r)\psi'''(r)}{r^6}\right.\\\nonumber&+&\left.\left.
\frac{112\zeta^2\psi'(r)}{r^8}-
\frac{448\zeta^2\psi(r)}{r^9}\frac{1568\zeta\psi^2(r)}{r^9}-\frac{168\zeta\psi'(r)}{r^8}
+\frac{672\zeta\psi(r)}{r^9}\right.\right.\\\nonumber&+&\left.\left.\frac{80\zeta\psi''(r)}{r^6}
+\frac{16\zeta^2\psi'^2(r)}{r^8}+\frac{16\zeta^2\psi(r)\psi''(r)}{r^8}
\frac{576\zeta^2\psi^2(r)}{r^10}\right)\right)/(A^2+\alpha\\\nonumber
&\times&B)^\frac{3}{2}))-\frac{1}{r^2}(2\psi'(r)
(-1+(\alpha(\frac{3\zeta}{r^3}(1-\frac{\psi(r)}{r})+\frac{2\zeta}{r^4}(1-\frac{\psi(r)}{r})
+\frac{2}{r^2}(1\\\nonumber&-&\frac{\psi(r)}{r})))/(A^2+\beta
B)^\frac{1}{2}))+((\frac{5\psi(r)}{r}-2-\frac{3\psi^2(r)}{r^2}
-3\psi'(r)(1-\frac{\psi(r)}{r}))\\\nonumber&\times&\alpha\beta^2C)/(r^3(A^2+\beta
B)^\frac{3}{2})+2(\frac{1}{r}
(2-\frac{\psi(r)}{r}-\frac{\psi'(r)}{r})(\frac{\zeta}{r^2}+\frac{1}{r})+2(1\\\nonumber
&-&\frac{\psi(r)}{r})(\frac{\zeta^2}{r^4}
-\frac{2\zeta}{r^3}))(-1+(\alpha A)/(A^2+\beta
B)^\frac{1}{2})-4\alpha(A^2+\beta B)^\frac{1}{2}\\\nonumber&+&A
(-1+(\alpha A)/(A^2+\beta B)^\frac{1}{2})+(\frac{2}{r^2}(1
-\frac{2\psi(r)}{r})+\frac{4\zeta}{r^3}(1-\frac{\psi(r)}{r}))(-1\\\nonumber&+&(\alpha
A)/(A^2+\beta B)^\frac{1}{2})
+\frac{1}{r}(4(1-\frac{\psi(r)}{r})(-(\alpha A^2)/(A^2+\alpha \beta
B)^\frac{3}{2}+\alpha\\\nonumber&/& (A^2+\beta
B)^\frac{1}{2})(-\frac{9\zeta}{r^4}\left(1-\frac{\psi(r)}{r}\right)+\frac{3\zeta}{r^3}
\left(-\frac{\psi'(r)}{r}+\frac{\psi(r)}{r^2}\right)
-\frac{8\zeta}{r^5}\left(1\right.\\\nonumber
&-&\left.\frac{\psi(r)}{r} \right)+\frac{2\zeta}{r^4}
\left(-\frac{\psi'(r)}{r}+\frac{\psi(r)}{r^2}\right)-\frac{4}{r^3}
\left(1-\frac{\psi(r)}{r}\right)+\frac{2}{r^2}
\left(-\frac{\psi'(r)}{r}\right.\\\nonumber&+&\left.
\frac{\psi(r)}{r^2}\right)))+(2B\alpha\beta)/
(A^2+\beta B)^ \frac{1}{2}+4\left(1-\frac{\psi(r)}{r}\right)
\left(\frac{\zeta}{r^2}+\frac{1}{r}\right)(-(\alpha\\\nonumber
&\times&A^2)/(A^2+\beta B)^\frac{3}{2} +\alpha/(A^2+\beta
B)^\frac{1}{2})
(-\frac{9\zeta}{r^4}\left(1-\frac{\psi(r)}{r}\right)+\frac{3\zeta}{r^3}
\left(\frac{\psi(r)}{r^2}\right.\\\nonumber
&-&\left.\frac{\psi'(r)}{r}\right)-\frac{8\zeta}{r^5}\left(1-\frac{\psi(r)}{r}\right)
+\frac{2\zeta}{r^4}
\left(-\frac{\psi'(r)}{r}+\frac{\psi(r)}{r^2}\right)
-\frac{4}{r^3}\left(1-\frac{\psi(r)}{r}\right)\\\nonumber&+&\frac{2}{r^2}
\left(-\frac{\psi'(r)}{r}+\frac{\psi(r)}{r^2}\right))
+\frac{12\zeta}{r^3}\left(1-\frac{\psi(r)}{r}\right)
+\frac{8\zeta}{r^4}\left(1-\frac{\psi(r)}{r}\right)=0.
\end{eqnarray}

\end{document}